\documentclass[fleqn,usenatbib]{mnras}

\usepackage{newtxtext,newtxmath}
\usepackage[T1]{fontenc}

\DeclareRobustCommand{\VAN}[3]{#2}
\let\VANthebibliography\thebibliography
\def\thebibliography{\DeclareRobustCommand{\VAN}[3]{##3}\VANthebibliography}

\usepackage{graphicx}	% Including figure files
\usepackage{amsmath}	% Advanced maths commands
\usepackage{cleveref}
\usepackage{gensymb}
\usepackage{threeparttable}
\usepackage{xcolor}
\usepackage{scalerel}
\usepackage{tikz}
\usetikzlibrary{svg.path}

\definecolor{orcidlogocol}{HTML}{A6CE39}
\tikzset{
  orcidlogo/.pic={
    \fill[orcidlogocol] svg{M256,128c0,70.7-57.3,128-128,128C57.3,256,0,198.7,0,128C0,57.3,57.3,0,128,0C198.7,0,256,57.3,256,128z};
    \fill[white] svg{M86.3,186.2H70.9V79.1h15.4v48.4V186.2z}
                 svg{M108.9,79.1h41.6c39.6,0,57,28.3,57,53.6c0,27.5-21.5,53.6-56.8,53.6h-41.8V79.1z M124.3,172.4h24.5c34.9,0,42.9-26.5,42.9-39.7c0-21.5-13.7-39.7-43.7-39.7h-23.7V172.4z}
                 svg{M88.7,56.8c0,5.5-4.5,10.1-10.1,10.1c-5.6,0-10.1-4.6-10.1-10.1c0-5.6,4.5-10.1,10.1-10.1C84.2,46.7,88.7,51.3,88.7,56.8z};
  }
}

\newcommand\orcidicon[1]{\href{https://orcid.org/#1}{\mbox{\scalerel*{
\begin{tikzpicture}[yscale=-1,transform shape]
\pic{orcidlogo};
\end{tikzpicture}
}{|}}}}

\usepackage{hyperref}

\title[SN 2021gno: Ca-rich transient]{SN 2021gno: a Calcium-rich transient with double-peaked light curves}

\author[K. Ertini et al.]{
K. Ertini,$^{\orcidicon{0000-0001-7251-8368},1,2}$\thanks{E-mail: keilaertini@fcaglp.unlp.edu.ar },
G. Folatelli,$^{1,2,3}$,
Martinez, L.,$^{1,2,4}$,
Bersten, M. C.,$^{1,2,3}$,
Anderson, J.~P.,$^{\orcidicon{0000-0003-0227-3451},5,6}$, 
Ashall, C.,$^{7}$,
\newauthor
Baron, E.,$^{8,9}$,
Bose, S.,$^{10,11}$,
Brown, P.~J.,$^{\orcidicon{0000-0001-6272-5507},12}$,
Burns, C.,$^{13}$,
DerKacy, J.~M.,$^{\orcidicon{0000-0002-7566-6080},8,7}$,
Ferrari, L.,$^{1,2}$,
\newauthor
Galbany, L.,$^{\orcidicon{0000-0002-1296-6887},14,15}$,
Hsiao, E.,$^{16}$,
Kumar, S.,$^{\orcidicon{0000-0001-8367-7591},16}$,
Lu, J.,$^{\orcidicon{0000-0002-3900-1452},16}$,
Mazzali, P.,$^{17,18}$,
Morrell, N.,$^{19}$,
Orellana, M.,$^{4,20}$,
\newauthor
Pessi, P.~J.,$^{\orcidicon{0000-0002-8041-8559},1}$,
Phillips, M.~M.,$^{\orcidicon{0000-0003-2734-0796},19}$,
Piro, A.~L.,$^{13}$,
Polin, A.,$^{13,21}$,
Shahbandeh, M.,$^{\orcidicon{0000-0002-9301-5302},16}$,
Shappee, B.~J.,$^{\orcidicon{0000-0003-4631-1149},22}$,
\newauthor
Stritzinger, M.,$^{23}$,
Suntzeff, N.~B.,$^{12}$,
Tucker, M.,$^{22}$,
Elias-Rosa, N.,$^{24,14}$,
Kuncarayakti, H.,$^{25,26}$,
\newauthor
Guti\'errez, C.~P.,$^{\orcidicon{0000-0003-2375-2064},25,26}$,
Kozyreva, A.,$^{27}$,
M\"uller-Bravo, T.~E.,$^{\orcidicon{0000-0003-3939-7167},14,15}$,
Chen, T.~-W.,$^{\orcidicon{0000-0002-1066-6098},28}$,
Hinkle, J.~T.,$^{22}$,
\newauthor
Payne, A.~V.,$^{\orcidicon{0000-0003-3490-3243},22}$,
Sz{\'e}kely, P.,$^{29,30}$,
Szalai, T.,$^{29,30,31}$,
Barna, B.,$^{29}$,
K{\"o}nyves-T{\'o}th, R.,$^{31,33,34}$,
B{\'a}nhidi, D.,$^{32}$,
\newauthor
B{\'i}r{\'o}, I.~B.,$^{30,32}$,
Cs{\'a}nyi, I.,$^{32}$,
Kriskovits, L.,$^{31,33}$, 
P{\'a}l, A.,$^{31,33,35,36,37}$,
Szab{\'o}, Zs.,$^{31,33,38,39}$, 
Szak{\'a}ts, R.,$^{31,33}$, 
\newauthor
Vida, K.,$^{31,33}$, 
Vink{\'o} J.,$^{31,33,35,40}$,
Gromadzki, M.,$^{\orcidicon{0000-0002-1650-1518},41}$,
Harvey, L.,$^{42}$,
Nicholl, M.,$^{\orcidicon{0000-0002-2555-3192},43}$,
Paraskeva, E.,$^{44}$,
\newauthor
Young, D.~R.,$^{\orcidicon{0000-0002-1229-2499},45}$,
Englert, B.,$^{46}$
\\
$^{1}$Facultad de Ciencias Astronómicas y Geofísicas, Universidad Nacional de La Plata, Paseo del Bosque S/N, B1900FWA, La Plata, Argentina\\
$^{2}$Instituto de Astrofísica de La Plata (IALP), CCT-CONICET-UNLP, Paseo del Bosque S/N, B1900FWA, La Plata, Argentina\\
$^{3}$Kavli Institute for the Physics and Mathematics of the Universe (WPI), The University of Tokyo, 5-1-5 Kashiwanoha, Kashiwa, Chiba, 277-8583, Japan\\
$^{4}$Universidad Nacional de R\'io Negro. Sede Andina, Mitre 630 (8400) Bariloche, Argentina\\
$^{5}$European Southern Observatory, Alonso de C\'ordova 3107, Casilla 19, Santiago, Chile\\
$^{6}$Millennium Institute of Astrophysics MAS, Nuncio Monsenor Sotero Sanz 100, Off. 104, Providencia, Santiago, Chile\\
$^{7}$Department of Physics, Virginia Tech, Blacksburg, VA 24061, USA\\
$^{8}$Homer L. Dodge Department of Physics and Astronomy, University of Oklahoma, 440 W. Brooks, Norman, OK 73019-2061, USA\\
$^{9}$Hamburger Sternwarte, Gojenbergsweg 112, 21029 Hamburg, Germany\\
$^{10}$Department of Astronomy, The Ohio State University, 140 W. 18th Avenue, Columbus, OH 43210, USA\\
$^{11}$Center for Cosmology and AstroParticle Physics (CCAPP), The Ohio State University, 191 W. Woodruff Avenue, Columbus, OH 43210, USA\\
$^{12}$Mitchell Institute for Fundamental Physics and Astronomy, Department of Physics and Astronomy, Texas A\&M University 4242 TAMU, College Station, TX, USA\\
$^{13}$The Observatories of the Carnegie Institution for Science, 813 Santa Barbara St., Pasadena, CA 91101, USA\\
$^{14}$Institute of Space Sciences (ICE, CSIC), Campus UAB, Carrer de Can Magrans, s/n, E-08193 Barcelona, Spain\\
$^{15}$Institut d’Estudis Espacials de Catalunya (IEEC), E-08034 Barcelona, Spain\\
$^{16}$Department of Physics, Florida State University, 77 Chieftain Way, Tallahassee, FL, 32306, USA\\
$^{17}$Astrophysics Research Institute, Liverpool John Moores University, IC2, Liverpool Science Park, 146 Brownlow Hill, Liverpool L3 5RF,UK\\
$^{18}$Max-Planck-Institut f\"ur Astrophysik, Karl-Schwarzschild Str. 1, D-85748 Garching, Germany\\
$^{19}$Carnegie Observatories, Las Campanas Observatory, Casilla 601, La Serena, Chile\\
$^{20}$Consejo Nacional de Investigaciones Científicas y Técnicas (CONICET), Argentina\\
$^{21}$TAPIR, Walter Burke Institute for Theoretical Physics, 350-17, Caltech, Pasadena, CA 91125, USA\\
$^{22}$Institute for Astronomy, University of Hawai\`{}i at Manoa, 2680 Woodlawn Dr., Honolulu, HI 96822\\
$^{23}$Department of Physics and Astronomy, Aarhus University, Ny Munkegade 120, DK-8000 Aarhus C, Denmark\\  
$^{24}$INAF – Osservatorio Astronomico di Padova, Vicolo dell’Osservatorio 5, I-35122 Padova, Italy\\
$^{25}$Tuorla Observatory, Department of Physics and Astronomy, FI-20014 University of Turku, Finland\\
$^{26}$Finnish Centre for Astronomy with ESO (FINCA), FI-20014 University of Turku, Finland\\
$^{27}$Heidelberger Institut f \"ur Theoretische Studien, Schloss-Wolfsbrunnenweg 35, 69118 Heidelberg, Germany\\
$^{28}$The Oskar Klein Centre, Department of Astronomy, Stockholm University, AlbaNova, SE-10691 Stockholm, Sweden\\
$^{29}$Department of Experimental Physics, Institute of Physics, University of Szeged, H-6720 Szeged, D{\'o}m t{\'e}r 9, Hungary \\
$^{30}$ELKH-SZTE Stellar Astrophysics Research Group, H-6500 Baja, Szegedi {\'u}t, Kt. 766, Hungary \\
$^{31}$Konkoly Observatory, Research Centre for Astronomy and Earth Sciences, H-1121 Budapest, Konkoly Thege Mikl{\'o}s {\'u}t 15-17, Hungary \\
$^{32}$Baja Astronomical Observatory of University of Szeged, H-6500 Baja, Szegedi {\'u}t, Kt. 766, Hungary \\
$^{33}$CSFK, MTA Centre of Excellence, Budapest, Konkoly Thege Mikl{\'o}s {\'u}t 15-17, Hungary \\
$^{34}$ELTE E{\"o}tv{\"o}s Lor{\'a}nd University, Gothard Astrophysical Observatory, Szombathely, Hungary \\
$^{35}$E{\"o}tv{\"o}s Lor{\'a}nd University, Department of Astronomy, P{\'a}zm{\'a}ny P{\'e}ter s{\'e}t{\'a}ny 1/A, 1117 Budapest, Hungary \\
$^{36}$E{\"o}tv{\"o}s Lor{\'a}nd University, Institute of Physics, P{\'a}zm{\'a}ny P{\'e}ter s{\'e}t{\'a}ny 1/A, 1117 Budapest, Hungary \\
$^{37}$MIT Kavli Institute for Astrophysics and Space Research, 70 Vassar Street, Cambridge, MA 02109, USA \\
$^{38}$Max-Planck-Institut f{\"u}r Radioastronomie, Auf dem Hügel 69, 53121 Bonn, Germany \\
$^{39}$Scottish Universities Physics Alliance (SUPA), School of Physics and Astronomy, University of St Andrews, North Haugh, St Andrews, KY16 9SS, UK \\
$^{40}$Department of Optics \& Quantum Electronics, Institute of Physics, University of Szeged, H-6720 Szeged, D{\'o}m t{\'e}r 9, Hungary\\
$^{41}$Astronomical Observatory, University of Warsaw, Al. Ujazdowskie 4, 00-478 Warszawa, Poland\\
$^{42}$School of Physics, Trinity College Dublin, The University of Dublin, Dublin 2, Ireland\\
$^{43}$Birmingham Institute for Gravitational Wave Astronomy and School of Physics and Astronomy, University of Birmingham, Birmingham B15 2TT, UK \\
$^{44}$Department of Physics and Astronomy, University of California, Davis, One Shields Avenue, CA 95616\\
$^{45}$Astrophysics Research Centre, School of Mathematics and Physics, Queen’s University Belfast, Belfast BT7 1NN, UK \\
$^{46}$Combate de los Pozos 1028, C1222AAL, Ciudad Autónoma de Buenos Aires, Argentina
}

\date{Accepted 2023 September 01. Received 2023 August 25; in original form 2023 January 05}

\pubyear{2021}

\begin{document}
\label{firstpage}
\pagerange{\pageref{firstpage}--\pageref{lastpage}}
\maketitle

\begin{abstract}
We present extensive ultraviolet (UV) and optical photometric and optical spectroscopic follow-up of supernova (SN)~2021gno by the "Precision Observations of Infant Supernova Explosions" (POISE) project, starting less than two days after the explosion. Given its
intermediate luminosity, fast photometric evolution, and quick transition to the nebular phase with spectra dominated by [Ca~II] lines, SN~2021gno belongs to the small family of Calcium-rich transients. Moreover, it shows double-peaked light curves, a phenomenon shared with only four other Calcium-rich events. The projected distance from the center of the host galaxy is not as large as other objects in this family. The initial optical light-curve peaks coincide with a very quick decline of the UV flux, indicating a fast initial cooling phase. Through hydrodynamical modelling of the bolometric light curve and line velocity evolution, we found that the observations are compatible with the explosion of a highly-stripped massive star with an ejecta mass of $0.8\,M_\odot$ and a $^{56}$Ni mass of $0.024~M_{\odot}$. 
The initial cooling phase (first light curve peak) is explained by the presence of an extended circumstellar material comprising $\sim$$10^{-2}\,M_{\odot}$ with an extension of $1100\,R_{\odot}$. We discuss if hydrogen features are present in both maximum-light and nebular spectra, and its implications in terms of the proposed progenitor scenarios for Calcium-rich transients. 

\end{abstract}


%%%%%%%%%%%%%%%%%%%%%%%%%%%%%%%%%%%%%%%%%%%%%%%%%%

%%%%%%%%%%%%%%%%% BODY OF PAPER %%%%%%%%%%%%%%%%%%

\section{Introduction}
In the past decade, the advent of high-cadence, all-sky surveys has drastically increased the number of discovered transients, including supernovae (SNe) which lie outside the traditional classification schemes. \citet[]{filippenko03} first classified "Calcium-rich" (or Ca-rich) SNe after observing four SNe with very strong emission from the Ca~II near-infrared triplet, and the [Ca~II] doublet near 730 nm. These SNe were first classified as Type Ib or Ic according to their early spectra, but they started to show distinct features as they evolved. \citet[]{perets10} studied a SN with similar properties, SN~2005E, which additionally was sub-luminous (absolute $B$-band peak magnitude of -14.8 mag) and rapidly evolving. 

As more of these transients have been discovered, the "Ca-rich" classification has become more robust. \citet[]{kasliwal12} identified the main properties of these transients as: peak luminosity values that are intermediate between those of novae and SNe ($-14$ to $-16.5$ mag), faster photometric evolution than normal SNe, photospheric velocities comparable to normal SNe, an early transition to the nebular phase, and nebular spectra dominated by calcium emission. These events have also been referred to as "Calcium-rich gap transients" because of their location in the luminosity gap between novae and SNe. Their name is reinforced, not only by strong calcium emission but by the high nebular Ca to O line ratio they exhibit. \citet[]{shen19} argued that while these transients have such high Ca to O line ratios, this does not necessarily imply a large production of Ca, thus they adopt the name "Calcium-strong Transients." 
Furthermore, \citet[]{polin21} showed that models with only 1$\%$ of Ca, produce nebular spectra that cool primarly through [Ca II] emission. For simplicity, in this work, we will refer to this class of events as "Ca-rich SNe". 

The Ca-rich class is relatively new, with few members, and they present several characteristics that make them a heterogeneous group. The majority of these transients are generally found in elliptical or S0 galaxies, which suggests a relationship with an old stellar population \citep[]{dong22}. Additionally, they are often found at large offsets from the nuclei of their host galaxies \citep[]{kasliwal12,lyman16}. \citet[]{foley15} studied a sample of thirteen Ca-rich transients and found that approximately one third of them were located at projected distances greater than 20~kpc from the galaxy nuclei. However, there is a fraction of these transients that are found well within their host galaxies and near star-forming regions, as is typical for core-collapse SNe. For example, iPTF16hgs \citep[]{de2018} was located at about 6~kpc (projected distance) from the center of its star-forming host galaxy, iPTF15eqv \citep[]{milisavljevic17} was offset by $6.5$~kpc from the center of its spiral host galaxy, and SN 2019ehk \citep[]{jacobsongalan20}, the closest Ca-rich transient to date, was found in the well-known spiral galaxy NGC 4321 (Messier 100) at $1.8$~kpc of projected distance from its nucleus. 

Adding to the heterogeneity within this class, there is a small number of Ca-rich transients that present double-peaked light curves. This is the case of iPTF16hgs, SN~2018lqo \citep[]{de2020}, and SN~2019ehk. There is another object, iPTF14gqr \citep[]{de2018b} that has a double-peaked light curve and shares all the photometric properties of a Ca-rich transient, but its spectrum at maximum light is quite distinct, and is thus not considered to be a canonical class member.

The diversity in the observed properties of these transients may suggest different physical origins. The origin of Ca-rich transients is still a matter of active debate, and multiple scenarios have been proposed \citep[see][for a recent review]{shen19}. Given the remote locations where these transients are usually found, the proposed progenitor systems frequently involve white dwarfs (WDs). For example, a WD passing close enough to an intermediate-mass black hole (IMBH; $M_{BH}$ $< 10^{5}$ $M_{\odot}$), or a WD tidally disrupted by a neutron star (NS) \citep[]{sell15}. However, many Ca-rich transients are explained as the detonation of a He shell on the surface of a WD \citep[]{perets10,wadman11}, which could be accreted from a He WD or a He star. There is, however, another explanation for the origin of a minority of Ca-rich transients that are found relatively close to their host-galaxy centers and near star-forming regions. They could be explained by the core-collapse (CC) explosion of a massive star, which has been stripped of most of its hydrogen envelope \citep[]{kawabata10}. Another possibility within the CC scenario is a He star in a binary system with a NS, which leads to the complete stripping of the He envelope, leading to what is called an ultra-stripped SN (USSN; \citealp[]{tauris13,tauris15}).

\citet[]{nakaoka21} suggested that the double-peaked iPTF16hgs, iPTF14gqr, and SN~2019ehk (and potentially iPTF15eqv, which has no pre-maximum observations) belong to a sub-population within the Ca-rich class that is associated with the USSN scenario. At the same time, \citet[]{de2021} suggested that iPTF15eqv and SN~2019ehk belong to a class of CCSNe with low-mass CO cores distinct from the thermonuclear Ca-rich transients found in old environments.  

In this paper we present photometry and spectra of SN~2021gno, a SN which is located in the relatively inner regions of its host galaxy NGC~4165. SN~2021gno was initially classified as a Type II SN \citep[]{hung21}, and later re-classified as Type Ib and Type IIb \citep[][respectively]{dahiwale21,perley21}, but its spectral as well as photometric evolution indicates it belongs to the Ca-rich class. We present observations obtained by the Precision Observations of Infant Supernova Explosions (POISE\footnote{\url{https://poise.obs.carnegiescience.edu/}}; \citealp[]{burns2021}) collaboration, starting $0.8$ days after discovery and spanning 25 days with $\approx$1 observation per day in all BV$ugri$-bands. Such an early and high cadence follow-up allows to observe initial light-curve peaks, which have been only seen in a handful of Ca-rich objects. Recently, \citet[]{jacobsongalan22} presented multi-wavelength  observations of SN~2021gno and suggested it was produced by the explosion of a WD star, likely due to the merger of a hybrid + CO WD system. Here we present an alternative progenitor scenario for this event: the CC explosion of a highly-stripped, massive star.

The paper is organised as follows. In Section~\ref{sec:obs}, we present the observations and data reduction of SN~2021gno. We analyse its photometric and spectroscopic properties in Section~\ref{sec:props}. In Section~\ref{sec:bolmod} we present the bolometric light curve and associated hydrodynamical modelling for our proposed progenitor scenario. Finally, in Section~\ref{sec:disc_concl} we provide a summary of our results and a discussion regarding the possible progenitor systems of Ca-rich transients.

\section{Observational data}
\label{sec:obs}
\begin{figure}
	\includegraphics[width=\columnwidth]{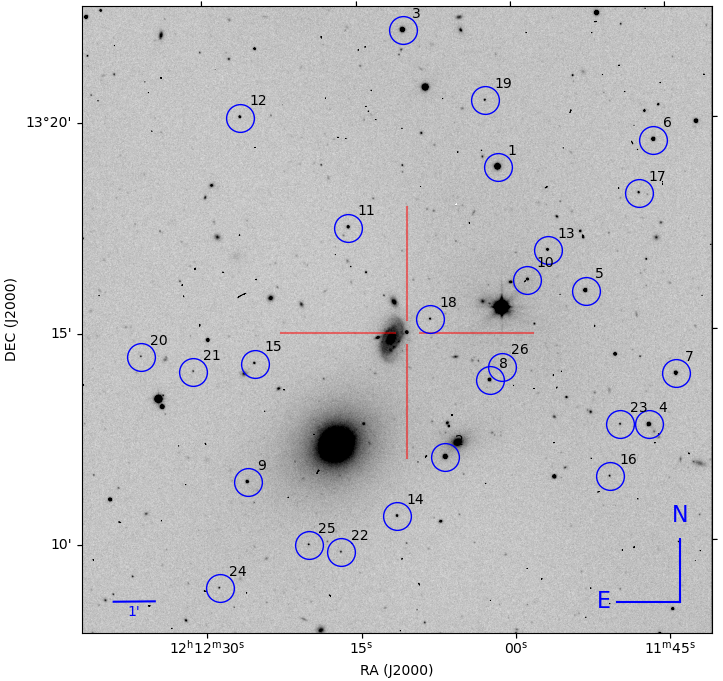}
    \caption{Swope Telescope $r$-band image of SN~2021gno (red cross) in NGC~4165. The blue circles indicate the local sequence stars. North is up and East is to the left.}
    \label{fig:2021gno}
\end{figure}

SN 2021gno (Figure~\ref{fig:2021gno}) was discovered by the Zwicky Transient Facility (ZTF; \citealt[]{masci19}) on UTC 2021 Mar 20 05:38:54.24 ($\mathrm{JD}=2459293.73$) with a magnitude of $r=18.2$ mag. Its coordinates are $\alpha = 12^{h}12^{m}10^{s}.294$ and $\delta =+13 \degree 14'57''.03$. The object was located in the galaxy NGC~4165, at an angular distance of $24.54''$ from its center, which corresponds to a projected physical separation of $4.56$~kpc. NGC~4165 has a reported heliocentric redshift of $z = 0.0062$ \citep[]{albareti17}. A distance of $38.4\pm1.8$~Mpc was adopted for NGC~4165 (see Section \ref{sec:dist_ext} for more details).

The last non-detection was reported by ZTF on $\mathrm{JD} = 2459291.85$, less than two days before the detection, with a limiting magnitude of $r=20.54$ mag. We take the estimated explosion epoch as the midpoint between the last non-detection and the first detection with an uncertainty equal to half the interval between those epochs, therefore at $\mathrm{JD}=2459292.79 \pm 0.94$. Unless noted otherwise, we will provide all epochs relative to this adopted explosion date in the rest frame of the SN.

\subsection{Photometry}
\label{sec:phot} 
The POISE collaboration started its follow-up observations of SN~2021gno shortly after its discovery. Optical photometry in the $BVugri$ bands was obtained with the 1.0~m Swope Telescope at Las Campanas Observatory (LCO), Chile. The observations were reduced using the POISE photometric pipeline, which closely follows the Carnegie Supernova Project (CSP) pipeline outlined in \citet[]{contreras10} and \citet[]{krisciunas17}. Briefly, the raw data are bias and flat-field corrected based on nightly calibrations. A world coordinate system (WCS) plate solution is computed using the Refcat2 catalog of standards \citep[]{tonry18}. Being a follow-up program rather than a discovery survey, POISE does not have reference images with which to do host galaxy subtractions. Instead, public data from Pan-STARRS \citep[]{kaiser02} or SkyMapper \citep[]{wolf18} are used for preliminary photometry. In the case of SN~2021gno, the host galaxy light is negligible and mismatches between these survey filters and our Swope filters do not introduce significant errors. 
Photometric calibration is done by observing \citet[]{landolt92} standards for $BV$ and \citet[]{smith02} standards for $ugri$. Using these standards, we computed colour terms that transform the instrumental magnitudes to the standard systems (see \citealt[]{krisciunas17} for the colour term equations and coefficients as well as the atmospheric extinction corrections). After correcting for atmospheric extinction, the colour terms are used in reverse to transform the \citet[]{landolt92} and \citet[]{smith02} standard magnitudes into Swope natural system magnitudes. These natural magnitudes are then used to calibrate local sequence stars in the field of SN~2021gno (see Figure~\ref{fig:2021gno}). Finally, the calibrated natural magnitudes of the local sequence stars are used to calibrate the relative photometry of SN~2021gno as a function of time. The net result is that the POISE light-curves are in the CSP natural system for the Swope telescope. Optical photometry in the Swope natural system is listed in Table~\ref{tab:phot_opt}. 

Once POISE observations were finished, we followed SN~2021gno as part of the Aarhus-Barcelona cosmic FLOWS project\footnote{\url{https://flows.phys.au.dk/}} using the Las Cumbres Observatory Global Telescope (LCOGT; \citealp[]{brown13}) network of 1~m telescopes equipped with the Sinistro cameras and the $BVgri$ filters. These data were calibrated using the $BVgri$ local sequence magnitudes in the CSP natural system from the Swope observations. LCOGT data is listed in Table~\ref{tab:phot_opt_lcogt}. 

SN~2021gno was also observed in the $BgVriz$ bands with two identical $0.8$~m telescopes at the Baja Observatory and Konkoly Observatory, both located in Hungary. Image subtraction was performed using a late-time template image obtained at $\mathrm{JD}=2459586.6$ (i.e.\ 293.81 days after the explosion) and the photometric calibration was based on field stars in the Pan-STARRS DR1 catalog \citep[]{chambers16}. Photometry from Baja and Konkoly Observatories is listed in Table~\ref{tab:phot_opt_bajakonk}. 

Additionally, space-based observations were triggered with the Ultraviolet/Optical Telescope (UVOT; \citealp[]{roming05}) at the Neil Gehrels Swift Observatory \citep[]{gehrels04}, starting on 2021 Mar 20 (JD = 2459294.044). The images were extracted from NASA's High Energy Astrophyisics Science Archive Research Center (HEASARC\footnote{\url{https://heasarc.gsfc.nasa.gov/}}), in $w2$, $m2$, $w1$, $u$, $b$, and $v$ filters. Aperture photometry was performed following the procedures in \citet{brown09}, with a $3''$ aperture, subtracting the galaxy count rate measured in a $3''$ aperture on observations from 2021 Jul 19. The magnitudes were calculated by applying a re-computed aperture correction and using the zero points from \citet{breeveld11} in the Swift system. The final Vega magnitudes are listed in Table~\ref{tab:phot_uv}. The Neil Gehrels Swift Observatory also triggered the X-Ray Telescope \citep[]{burrows05} and detected a bright X-ray emission at $\approx$0.3~days after the discovery \citep[]{jacobsongalan22}. An analysis of the X-ray observations is beyond the scope of this paper, but see \citealt[]{jacobsongalan22} for more details.

\begin{figure*}
	\includegraphics[scale=1]{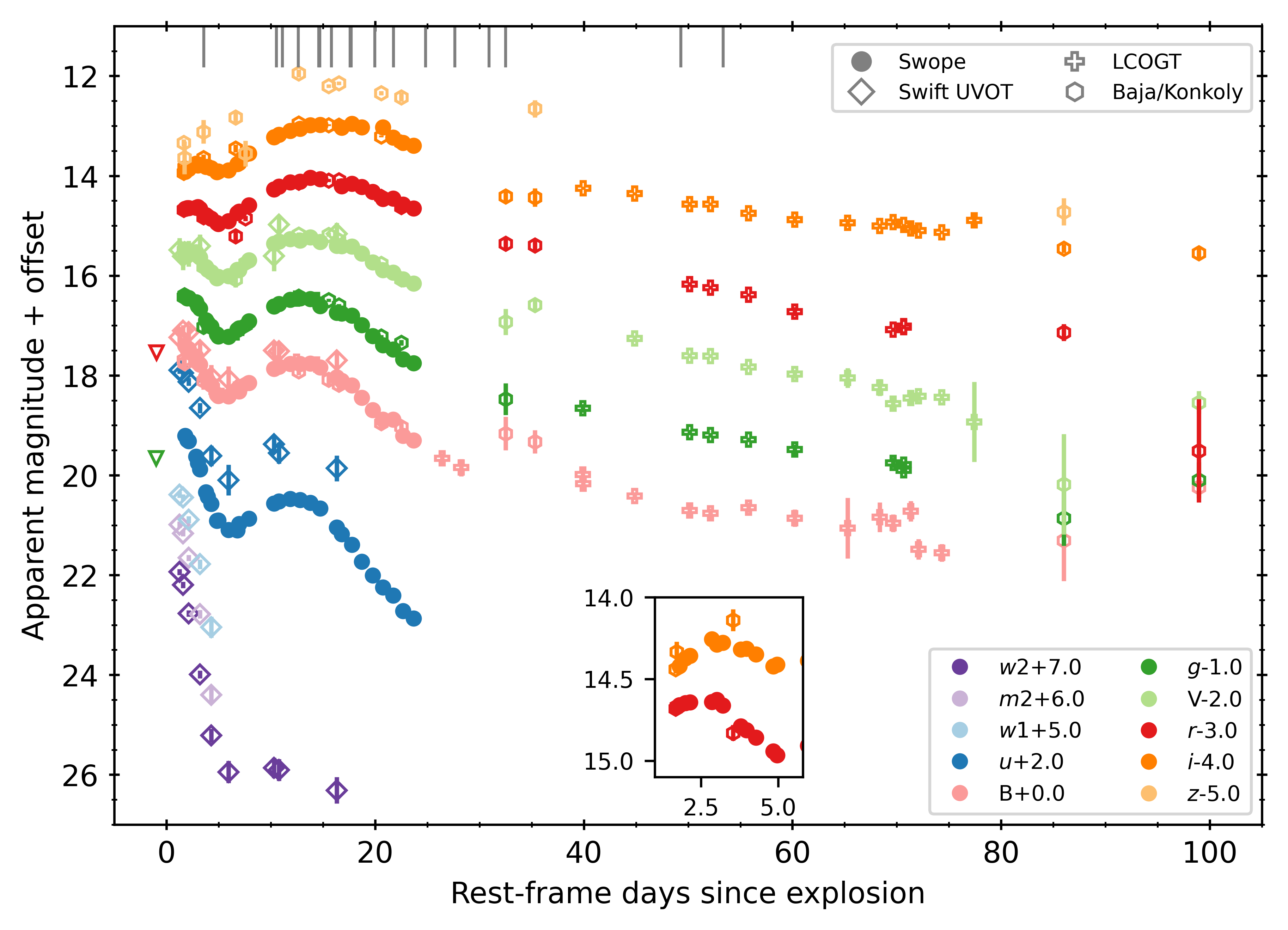}
    \caption{Optical and UV light curves of SN~2021gno, corrected for Galactic extinction. For clarity, the light curves are shifted by the offsets indicated in the lower legend. Different instruments are indicated as different markers. Filled symbols are used for Swope photometry and open symbols for other telescopes. Although they are plotted with the same colours, note that the transmission functions of the Swift $ubv$ bands, specially $u$, do not match those of the ground-based $uBV$ bands. ZTF non-detections are indicated as triangles. The inset plot shows the period around the first peak of the light curve, in $r$ and $i$ bands. Rest-frame epochs of optical spectra are marked as gray lines along the top axis.}
    \label{fig:lcs}
\end{figure*}

The resulting UV and optical light curves of SN 2021gno are shown in Figure \ref{fig:lcs}. 

\subsection{Spectroscopy}
Spectroscopic observations started soon after discovery, at approximately 4 days relative to the time of the explosion. The observations continued until 2021 Jul 17, covering 116 days. The log of spectroscopic observations is listed in Table \ref{tab:log_spec}. 

Three spectra were taken before the second maximum with the SPectrograph for the Rapid Acquisition of Transients (SPRAT; \citealp[]{piascik14}) on the 2.0~m Liverpool Telescope at Observatorio del Roque de los Muchachos, and with the Wide-Field Spectrograph (WiFeS; \citealp[]{dopita07}) on the Australian National University 2.3~m telescope at the Siding Spring Observatory. After maximum light, spectra were taken with the following telescopes/instruments: 1) the ESO Faint Object Spectrograph and Camera (EFOSC2; \citealp[]{buzzoni84}) on the 3.58~m New Technology Telescope (NTT) at La Silla Observatory, within the Public European Southern Observatory Spectroscopic Survey of Transient Objects (ePESSTO+) collaboration \citep[]{smartt15}; 2) the Alhambra Faint Object Spectrograph and Camera (ALFOSC) on the 2.5~m Nordic Optical Telescope (NOT) at Observatorio del Roque de los Muchachos, which were obtained in collaboration with the Nordic optical telescope Unbiased Transient Survey 2 (NUTS2\footnote{\url{https://nuts.sn.ie/}}); 3) the Supernova Integral Field Spectrogaph (SNIFS; \citealp[]{lantz04}) on the University of Hawaii 2.2~m Telescope (UH2.2m) at Mauna Kea, as a part of the SCAT survey \citep[]{tucker22}; 4) the Dual Imaging Spectrograph (DIS) on the 3.5~m Telescope at the Apache Point Observatory\footnote{Owned and operated by the Astrophysical Research Consortium.}; and 5) the Ohio State Multi-Object Spectrograph (OSMOS) on the 2.4~m Hiltner telescope at the MDM Observatory. Two nebular spectra were obtained on July 8 and 17, with the Optical System for Imaging and low-Intermediate-Resolution Integrated Spectroscopy (OSIRIS) on the Gran Telescopio Canarias at Observatorio del Roque de los Muchachos, and with the FOcal Reducer/low dispersion Spectrograph 2 (FORS2; \citealp[]{appenzeller98}), mounted on the Very Large Telescope at Paranal Observatory, as part of the FORS+ Survey of Supernovae in Late Times program (FOSSIL, Kuncarayakti
et al. in prep.). 

We reduced the EFOSC2 spectra with the PESSTO pipeline \citep[]{smartt15}. DIS spectra were reduced using IRAF \citep[]{tody86}, including standard methods for bias and flat-field corrections, with flux calibrations based on standard stars observed at a similar airmass and during the same night as the SN, and cosmic ray removal performed with via Lacosmic \citep[]{vandokkum01}. The 2D SNIFS frames were pre-processed and extracted into a 3D datacube using the methods of \citet[]{bacon01}. Then, a custom Python pipeline performed aperture photometry on the 3D datacube using a wavelength-dependent trace and the extracted spectra were placed on a relative flux scale using observations of spectrophotometric standard stars \citep[]{tucker22}. The OSMOS data were reduced with PyRAF-based \texttt{SimSpec}\footnote{\url{https://astro.subhashbose.com/simspec/}} pipeline. The SPRAT spectra were reduced using the standard liverpool telescope pipeline\footnote{\url{https://telescope.livjm.ac.uk/}}. The ALFOSC and GTC spectra were reduced using the dedicated pipeline FOSCGUI \footnote{FOSCGUI is a Python-based graphic user interface (GUI) developed by E. Cappellaro and aimed at extracting supernova spectroscopy and photometry obtained with FOSC-like instruments. A package description can be found at \url{http://sngroup.oapd.inaf.it/foscgui.html}}, which follows the standard procedures. This includes bias subtraction, flat-field correction, 1D extraction, and wavelength and flux calibration. The FORS2 spectrum was reduced using the ESOReflex \citep[]{freudlin13} pipeline following standard procedures, and using observations of spectrophotometric standard stars taken under the same grism setting.

\begin{figure*}
	\includegraphics[scale=0.85]{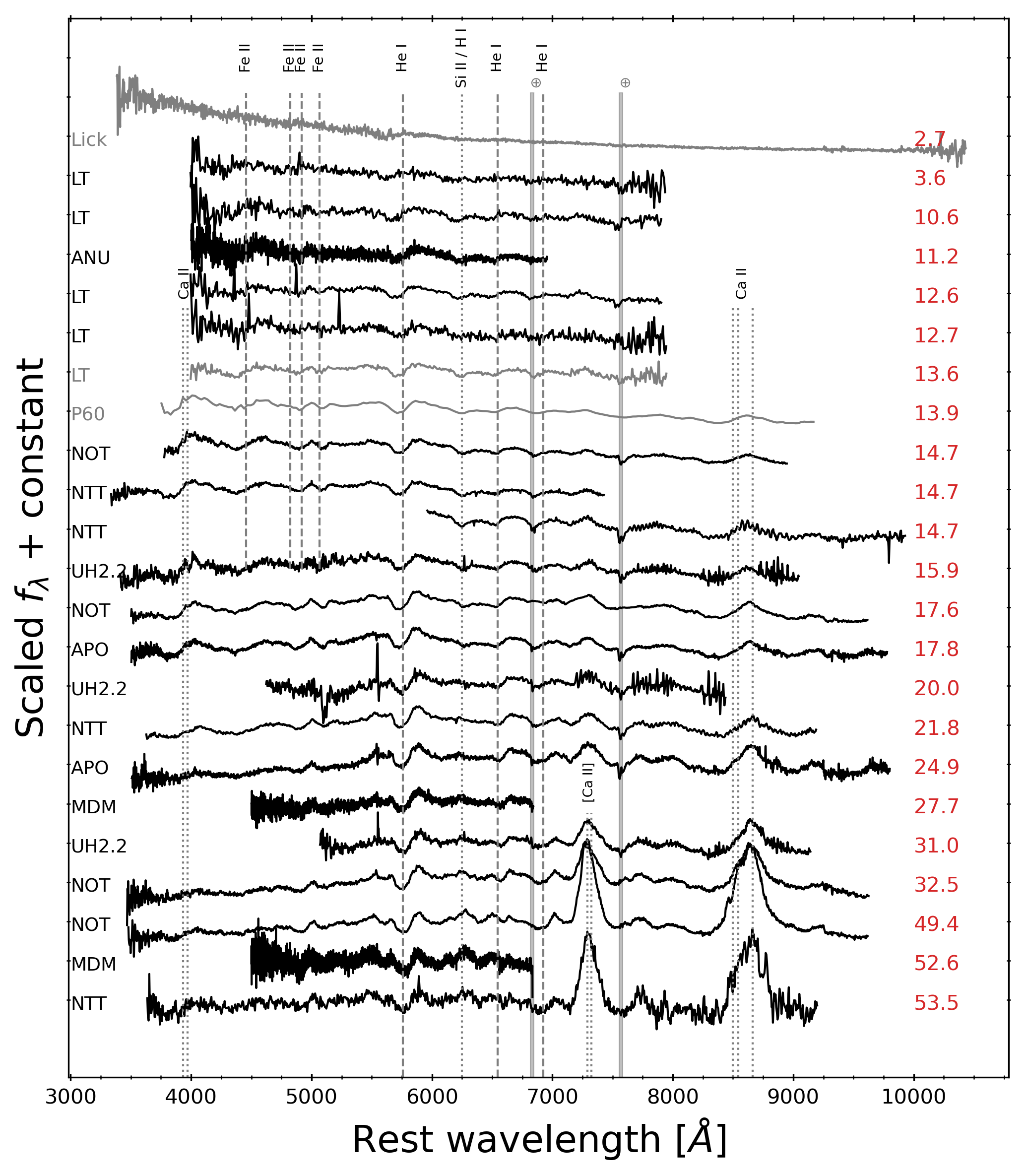}
    \caption{Spectroscopic evolution of SN 2021gno. Spectra taken by the POISE collaboration are plotted in black, and the three public spectra from TNS are plotted in gray. The phase in rest-frame days since explosion is given on the right-hand side of each spectrum, and the labels on the left-hand side indicate the telescope. Main optical lines at a fixed expansion velocity of 6000~km~s$^{-1}$ are marked with dashed gray vertical lines, while gray bands mark the location of telluric absorptions.}
    \label{fig:spec}
\end{figure*}

\begin{figure}
	\includegraphics[width=\columnwidth]{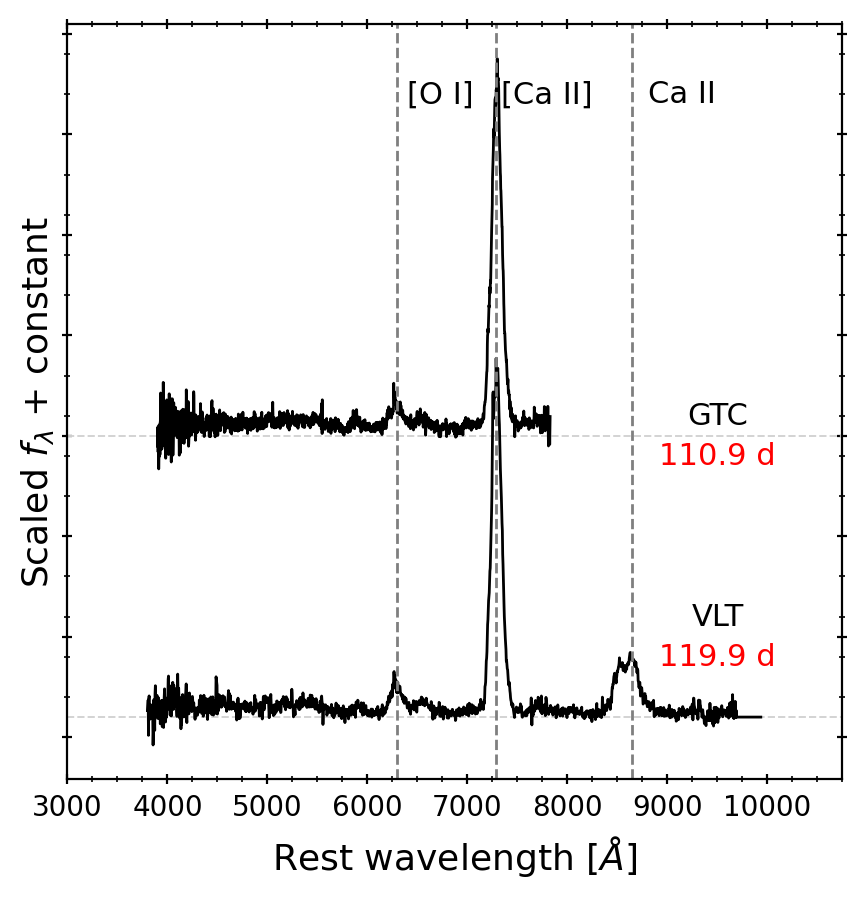}
    \caption{Nebular spectra of SN~2021gno. The labels on the right-hand side of each spectrum indicate the phase with respect to the explosion epoch (in red), and the telescope (in black). Main lines are marked in the rest frame with gray dashed vertical lines. Dashed horizontal lines mark zero flux.}
    \label{fig:nebspec}
\end{figure}

We have also included three public spectra of SN~2021gno available from the Weizmann Interactive Supernova Data Repository (WISeREP\footnote{\url{https://wiserep.weizmann.ac.il}}; \citealp[]{yaron12}). The spectral sequence of SN~2021gno is shown in Figure~\ref{fig:spec}. The two nebular spectra of SN~2021gno are shown in Figure~\ref{fig:nebspec}.

\begin{table}
	\caption{Log of spectroscopic observations. The phase is indicated in rest-frame days from explosion.}
	\label{tab:log_spec}
	\begin{tabular}{lcccc} 
		\hline
		Date & JD & Telescope & Instrument & Phase  \\
		\hline
		2021 Mar 22 & 2459296.44 & LT & SPRAT & 3.6 \\ 
		2021 Mar 29 & 2459303.437 & LT & SPRAT & 10.6  \\
		2021 Mar 30 & 2559304.071 & ANU 2.3-m & WiFes & 11.2  \\
		2021 Mar 31 & 2459305.493 & LT & SPRAT & 12.6  \\
		2021 Mar 31 & 2459305.540 & LT & SPRAT & 12.7  \\
        2021 Apr 03 & 2459307.568 & NOT & ALFOSC & 14.7  \\
		2021 Apr 03 & 2459307.572 & NTT & EFOSC & 14.7  \\
		2021 Apr 03 & 2459307.602 & NTT & EFOSC & 14.7  \\
		2021 Apr 04 & 2459308.763 & UH88 & SNIFS & 15.9  \\
		2021 Apr 05 & 2459310.489 & NOT & ALFOSC & 17.6  \\
		2021 Apr 06 & 2459310.655 & APO 3.5-m & DIS &  17.8 \\
		2021 Apr 08 & 2459312.917 & UH88 & SNIFS & 20.0  \\
		2021 Apr 10 & 2459314.714 & NTT & EFOSC & 21.8  \\
		2021 Apr 13 & 2459317.863 & APO 3.5-m & DIS & 24.8  \\
		2021 Apr 16 & 2459320.672 & MDM 2.4-m & OSMOS &  27.7 \\
		2021 Apr 19 & 2459323.953 & UH88 & SNIFS & 31.0  \\
		2021 Apr 21 & 2459325.539 & NOT & ALFOSC & 32.5  \\
		2021 May 06 & 2459342.479 & NOT & ALFOSC & 49.4  \\
		2021 May 11 & 2459345.718 & MDM 2.4-m & OSMOS & 52.6 \\
		2021 May 12 & 2459346.584 & NTT & EFOSC & 53.5  \\
		2021 Jul 08 & 2459404.425 & GTC & OSIRIS & 110.9  \\
		2021 Jul 17 & 2459413.478 & VLT & FORS2 &  119.9 \\ 
    	\hline
	\end{tabular}
\end{table}

\section{Properties}
\label{sec:props}
\subsection{Distance and extinction}
\label{sec:dist_ext}
The galaxy NGC~4165 also hosted the Type Ia SN~1971G, enabling a precise redshift-independent distance estimation. This was done by \citet[]{mueller94}, who derived a distance of $34.7$~Mpc using models of detonation, delayed detonation, and deflagration. In addition, \citet[]{hoeflich96} obtained a value of 36~Mpc using three different models of delayed detonation and one of deflagration. The distance was also estimated using the Tully-Fisher method by several authors, as listed in the NASA Extragalactic Database (NED\footnote{\url{https://ned.ipac.caltech.edu/}}). In particular, \citet[]{yasuda97} obtained a value of $57.5$ Mpc, and \citet[]{theureau07} obtained $57.6$~Mpc using $J$-band data, and $59.9$~Mpc using $H$-band data\footnote{By mistake the NED lists an additional distance from \citet[]{theureau07} with the note "mean", but this is a redshift-dependent measurement and should not be considered as a Tully-Fisher estimation.}. Additionally, NGC 4165 is possible member of a galaxy group that is included in the Cosmicflows-2 catalog \citep[]{tully13} whose brightest member is NGC 4168. \citet[]{tully13} give a distance to the group of $34.5\pm1.3$~Mpc (for $H_{0} = 74.4$~km~s$^{-1}$~Mpc$^{-1}$).

Given the large discrepancies among previous distance estimations, we recalculated it using the available $UBV$ photometry of SN~1971G \citep[][and references therein]{cadonau90} and the Type Ia SN light-curve fitting package SNooPy \citep[]{burns11}, with the \texttt{EBV2\_model} option. SNooPy adopts $H_{0} = 72$~km~s$^{-1}$~Mpc$^{-1}$, $\Omega_{M}=0.28$, and $\Omega_{\Lambda}=0.72$ \citep[]{spergel07}. Although there is pre-maximum $B$-band photometry (as early as $-17$ days), none of the observations cover the time of maximum light. This, along with the heterogeneous origin of the (pre-CCD) photometry, may be the reason why the SNooPy fits yield a negative $E(B-V)$ colour excess value. We considered such fits as unphysical and decided to fix the extinction to zero (consistent with the analysis below). With such a constraint the resulting distance modulus was $\mu=32.92\pm0.10$ mag, which corresponds to a distance of $38.4\pm1.8$~Mpc. That is the distance value that we adopt throughout this paper. Note that this distance is consistent with the Cosmicflows-2 catalog distance for the NGC4168 group.

Regarding the extinction, for the Milky-Way (MW) component we adopted a value of $E(B-V)=0.03$ mag from the infrared dust maps \citep[]{schlafly11} available from NED, and an extinction law from \citet[]{cardelli89} with $R_V=3.1$. To determine any possible host-galaxy extinction, we first examined the spectra for signs of Na~I~D absorption. We found no evidence of such an absorption at the redshift of the host galaxy in any of the available spectra. This suggests a smaller extinction component from the host than from the Milky Way. Thus we consider the host galaxy extinction to be negligible for the rest of the analysis.

\begin{figure*}
	\includegraphics[scale=0.42]{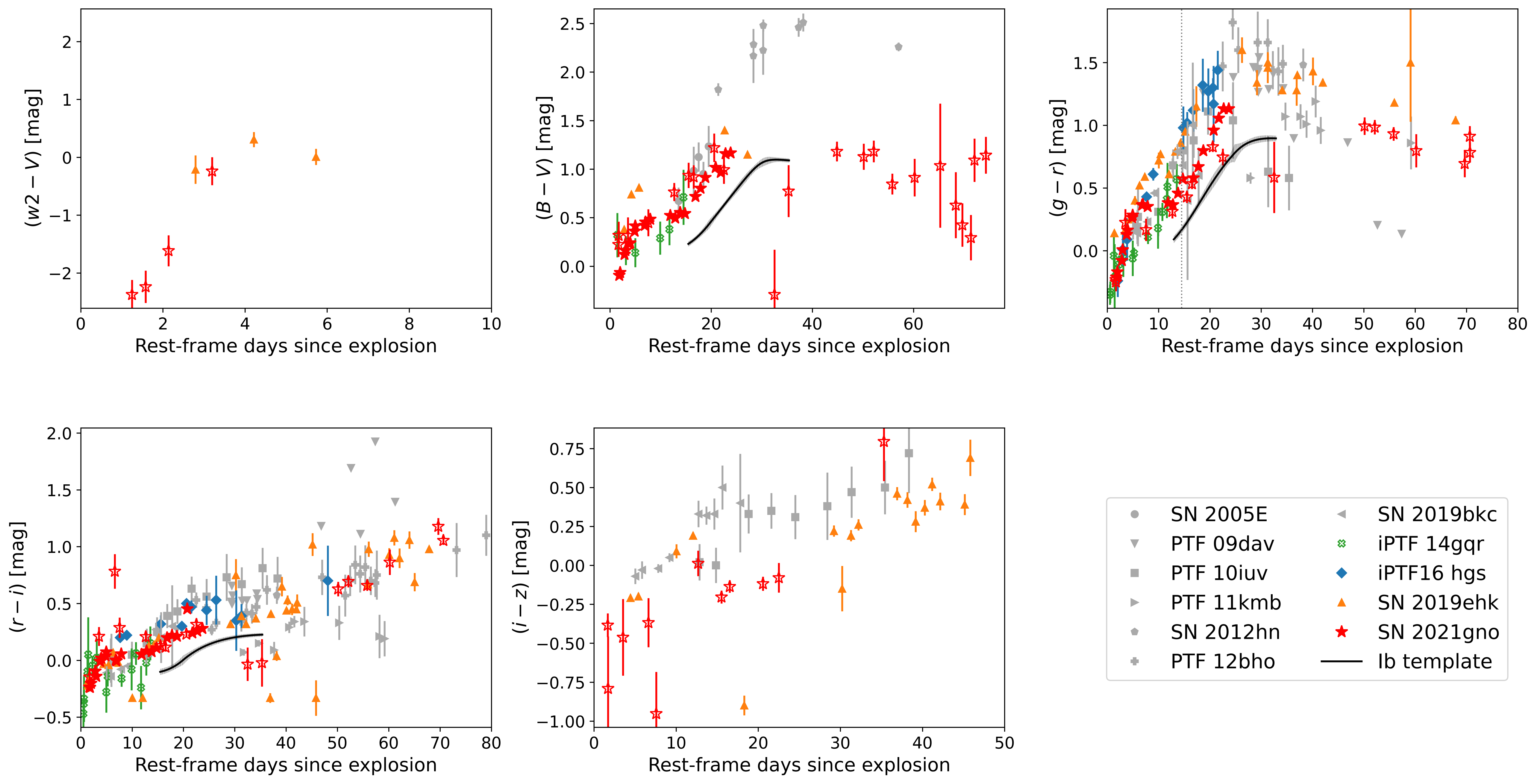}
    \caption{Colour curves of SN~2021gno (red stars) compared to other Ca-rich transients. Filled symbols are used for Swope photometry and open symbols for other telescopes. Ca-rich transients SN~2005E \citep[]{perets10}, PTF09dav \citep[]{sullivan11}, PTF10iuv \citep[]{kasliwal12}, PTF11kmb \citep[]{lunnan17}, SN~2012hn \citep[]{valenti14}, PTF12bho \citep[]{lunnan17}, SN~2016hnk \citep[]{jacobsongalan20a} and SN~2019bkc \citep[]{prentice20} are marked in gray. The double-peaked Ca-rich transients iPTF16hgs \citep[]{de2018}, SN~2019ehk \citep[]{jacobsongalan20}, and the peculiar double-peaked iPTF14gqr \citep[]{de2018b}, are marked in blue, orange, and green, respectively. For comparison, SN-Ib colour templates from \citet[]{stritzinger18} are shown as a solid black line. For reference, the epoch of $r$-band maximum light is marked with dotted lines in the (g-r) panel.}
    \label{fig:ex_comp}
\end{figure*}

\subsection{Light curves}
As seen in Figure \ref{fig:lcs}, SN~2021gno presents a double-peaked optical light curve. This property is shared with a small number of Ca-rich transients, the others being iPTF16ghs, SN~2018lqo, SN~2019ehk, and SN~2021inl \citep[]{de2018b,de2020,jacobsongalan20,jacobsongalan22}. The rise to the first peak in SN~2021gno is observed in the $r$ and $i$-bands, which is also seen in SN~2019ehk. In the rest of the bands the data show an initial decline, meaning that the first peak possibly occurred in those bands prior to the start of observations.

    We first analyzed the second peak, which we will call the main peak. We fit the light curves using a low-order polynomial in order to get peak magnitudes and rise times. Due to a gap in observations between 30 and 40 days after the explosion, we used observations taken before 30 days after the explosion for the fits, to avoid large uncertainties introduced by the lack of data. Also, to avoid dispersion in the fits, we excluded the $u$, $b$, and $v$ Swift bands from our analysis since their effective wavelengths are somewhat different from the standard ones \citep[]{poole08}. The results for all the bands are listed in Table~\ref{tab:lc_prop}. We obtained an absolute peak magnitude in the $r$-band of $M_{r}=-15.77\pm0.12$ mag at $\mathrm{JD}=2459307.25$, giving a rise time of $t_{\mathrm{r}}=14.46$ days. The uncertainty in the $r$-band absolute peak magnitude is the result of adding the uncertainties in the apparent peak magnitude, and  the distance in quadrature. For the $z$-band, these parameters were not calculated given the low cadence of the observations. We used the same interpolations of the $r$- and $i$-band light curves in order to estimate the magnitude and rise time of the first peak. The results are listed in the lower panel of Table \ref{tab:lc_prop}. Our estimated values are compatible with those given by \citet[]{jacobsongalan22}.

\begin{table}
	\caption{Light curve properties of SN~2021gno}
	\label{tab:lc_prop}
	\begin{threeparttable}
	\hspace*{-4cm}\begin{tabular}{p{0.03\textwidth}cp{0.03\textwidth}cp{0.045\textwidth}c} 
	    \hline
	    \multicolumn{6}{|c|}{Main peak} \\
		\hline
		Filter & $\mathrm{JD_{max}}$ & $t_\mathrm{r}$\tnote{a} & $m_\mathrm{max}$ & $M_\mathrm{max}$\tnote{b} & Decl. rate \\
		 & $-2459000$ & [days] & [mag] & [mag] & [mag d$^{-1}$] \\
		\hline
		$B$ & 305.59 & 12.7 & $17.92 \pm 0.09$ & $-14.99$ & $0.142 \pm 0.009$ \\
		$V$ & 306.51 & 13.6 & $17.32 \pm 0.05$ & $-15.60$ & $0.091 \pm 0.007$ \\
		$u$ & 305.08 & 12.2 & $18.62 \pm 0.04$ & $-14.29$ & $0.233 \pm 0.008$ \\
		$g$ & 305.92 & 13.0 & $17.56 \pm 0.09$ & $-15.35$ & $0.121 \pm 0.008$\\
		$r$ & 307.25 & 14.4 & $17.14 \pm 0.06$ & $-15.77$ & $0.073 \pm 0.005$\\
		$i$ & 307.72 & 14.8 & $17.03 \pm 0.05$ & $-15.88$ & $0.052 \pm 0.008$\\
    	\hline
         \multicolumn{6}{|c|}{First peak} \\
    	\hline
    	$w2$ & $\cdots$ & $\cdots$ & $\cdots$ & $\cdots$ & $1.08 \pm 0.02$ \\
    	$m2$ & $\cdots$ & $\cdots$ & $\cdots$ & $\cdots$ & $1.11 \pm 0.08$ \\
    	$w1$ & $\cdots$ & $\cdots$ & $\cdots$ & $\cdots$ & $0.88 \pm 0.07$ \\
    	$B$ & $\cdots$ & $\cdots$ & $\cdots$ & $\cdots$ &  $0.25 \pm 0.03$ \\
    	$V$ & $\cdots$ & $\cdots$ & $\cdots$ & $\cdots$ &  $0.16 \pm 0.01$ \\
    	$u$ & $\cdots$ & $\cdots$ & $\cdots$ & $\cdots$ &  $0.56 \pm 0.02$ \\
    	$g$ & $\cdots$ & $\cdots$ & $\cdots$ & $\cdots$ &  $0.25 \pm 0.02$ \\
    	$r$ & 295.30 & 2.5 & $17.70 \pm 0.01$ & $-15.21$ & $0.12 \pm 0.02$ \\
		$i$ & 295.50 & 2.7 & $17.80 \pm 0.01$ & $-15.11$ & $0.08 \pm 0.03$ \\
		\hline
	\end{tabular}
	\begin{tablenotes}
    \item[a] The uncertainty for all rise times is taken as $0.9$ days, such as the error in the explosion epoch, since it is the main source of error.
    \item[b] The uncertainty in $M_\mathrm{max}$ can be computed as the sum in quadrature of the errors in $m_\mathrm{max}$ and the estimated $0.1$ mag error in the distance.
    \end{tablenotes}
	\end{threeparttable}
\end{table}

We fit straight lines to the light curves (in magnitude scale) in order to obtain initial decline rates after the first peak in all bands. Using the data points between $2.3$ and 5 days from explosion, we obtained decline rates for the $r$ and $i$-band. Assuming the first peak in the other bands takes place before the observations start, we use the data points between the earliest observation and $4.5$ days after the explosion in order to fit decline rates for the rest of the bands (excluding the $z$-band). The results are given in Table~\ref{tab:lc_prop}. We note that the initial decline is slower the redder the band is. The decline rates after the main peak are also listed in Table~\ref{tab:lc_prop}. The same wavelength dependence as for the initial decline rates is observed.

The high cadence in observations of SN~2021gno across all bands allowed us to study its colour evolution compared with a comprehensive sample of Ca-rich transients taken from the literature. The extinction-corrected $(w2-V)$, $(B-V)$, $(g-r)$, $(r-i)$, and $(i-z)$ colour curves of SN~2021gno are shown in Figure~\ref{fig:ex_comp}. The comparison sample was constructed by taking all Ca-rich transients with available colours, and their photometry was corrected for Galactic extinction neglecting the host-galaxy component, except for SN~2019ehk, which has substantial extinction from its host \citep[]{nakaoka21,jacobsongalan20}. We adopted a value of $E(B-V)_{\mathrm{Host}}=0.47$ mag from \citet[]{jacobsongalan20} for this SN. The colour curves of SN~2021gno are broadly compatible with those of similar objects, although there is a substantial dispersion within the sample. Furthermore, SN~2021gno lies on the `blue edge' of the colour distribution, adding to the conclusion of a negligible host-galaxy extinction component (see Section~\ref{sec:dist_ext}).

SN~2021gno quickly evolved from blue to red colours during the first three weeks of evolution (through both peaks in the light curve). This suggests a rapid cooling of the ejecta (see Section~\ref{sec:bol}). The same behavior is observed for other double-peaked Ca-rich transients. Afterwards, the colours remained red and nearly constant for the rest of the evolution. As seen from the $(w2-V)$ panel in Figure~\ref{fig:ex_comp}, SN~2021gno is the first Ca-rich transient for which this colour is available right after the explosion. The existence of early UV $-$ optical colours is crucial to constrain the temperature during the fast initial evolution (see Section~\ref{sec:bol}).

As the majority of Ca-rich transients have spectra similar to type Ib SNe at maximum light, we also show in Figure \ref{fig:ex_comp} the Type Ib intrinsic colour templates of \citet[]{stritzinger18} for $(B-V)$, $(g-r)$ and $(r-i)$. The colour templates are given for times after maximum light, so we converted them to epochs after explosion using the rise times of SN~2021gno in the corresponding bands. Even if the overall photometric evolution of SN~2021gno is faster than that of typical stripped-envelope (SE)~SNe, the rate at which colours change after maximum light is comparable to those of Type Ib SNe, but with a nearly constant shift toward redder colours.

\begin{figure*}
	\includegraphics[scale=0.8]{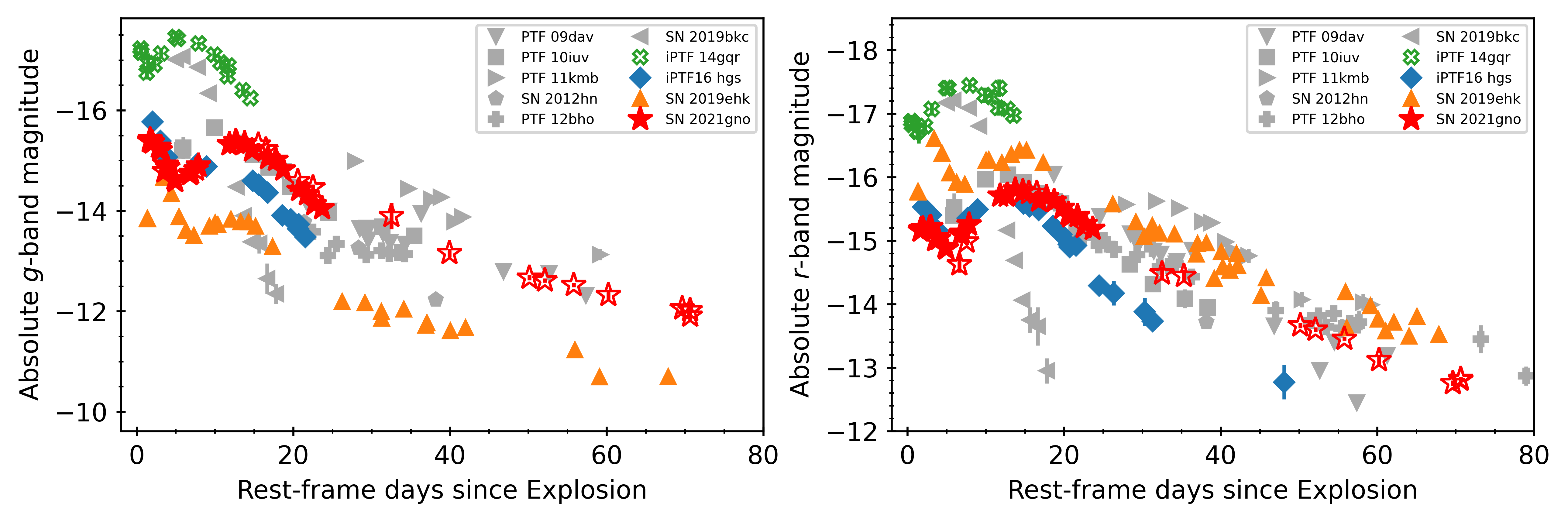}
    \caption{Absolute-magnitude light-curve comparison of SN~2021gno (red stars) and other Ca-rich transients ($g$-band on the left panel, and $r$-band on the right panel). Filled symbols are used for Swope photometry and open symbols for other telescopes. Double-peaked Ca-rich transients iPTF16hgs, SN~2019ehk, and the peculiar double-peaked iPTF14gqr are marked in blue, orange, and green, respectively, whereas objects with singly peaked light curves are shown in gray. References are listed in Figure~\ref{fig:ex_comp}.}
    \label{fig:r_lc_comp}
\end{figure*}

We compare the absolute $g$ and $r$-band light curve of SN~2021gno to that of several Ca-rich transients in Figure~\ref{fig:r_lc_comp}.
Similar to the colour evolution, the criteria for defining the comparison sample was to take the available photometry of all Ca-rich transients from the literature. The main peak of SN~2021gno is comparable in duration and luminosity with those of the other Ca-rich transients. The first peak is similar to those of iPTF16hgs and SN~2019ehk both in terms of duration and slope. The similarity of initial behavior, both in the light curve and in the colour evolution, among Ca-rich transients with double-peaked light curves may indicate a similar origin for this subgroup of events.

\subsection{Spectral evolution}
Figures~\ref{fig:spec} and \ref{fig:nebspec} show the spectral sequence of SN~2021gno, which was obtained between $2.7$ and 120 days relative to the time of the explosion. As is typical of Ca-rich SNe, SN~2021gno rapidly evolves to the nebular phase, with forbidden transitions that dominate at about 50 days after explosion (37 days after the main maximum). The last two spectra at 111 and 120 days are clearly nebular, with a very weak continuum and dominated by emission lines mostly from Ca~II. In the following sections we describe the properties of the spectra in the photospheric and nebular phases.

\subsubsection{Photospheric phase}
\label{sec:spec_photospheric}
After an initial nearly featureless spectrum with a blue continuum at $2.7$ days, the spectra of SN~2021gno become dominated by helium lines, most prominently He~I~$\lambda$5876 (Figure~\ref{fig:spec}). This continues until about 30 days past explosion. Although this feature could be due, at least in part, to Na~I~D, the identification as He~I is supported by the presence of an absorption that can be identified as He~I~$\lambda$6678. The He~I~$\lambda$7065 line may also be present, although it is affected by the telluric H$_2$O band. The identification is confirmed by our SYNOW analysis (see below). 

At times around maximum light, the spectrum resembles those of normal Type Ib SNe. Other spectral features can be identified as due to Ca~II~$\lambda\lambda\lambda$~8498,~8542,~8662 (the Ca~II IR triplet), a set of blended Fe~II lines where a weak Fe~II~$\lambda$5169 absorption is distinguishable, and possibly O~I~$\lambda$7774. An absorption is also present near 6250 \AA, at least until maximum light. This feature usually appears in Type Ib SNe and in Ca-rich transients as well, and it is often associated with Si~II~$\lambda$6355, but its identification is ambiguous \citep[e.g., see][]{folatelli14}.

\begin{figure}
	\includegraphics[width=\columnwidth]{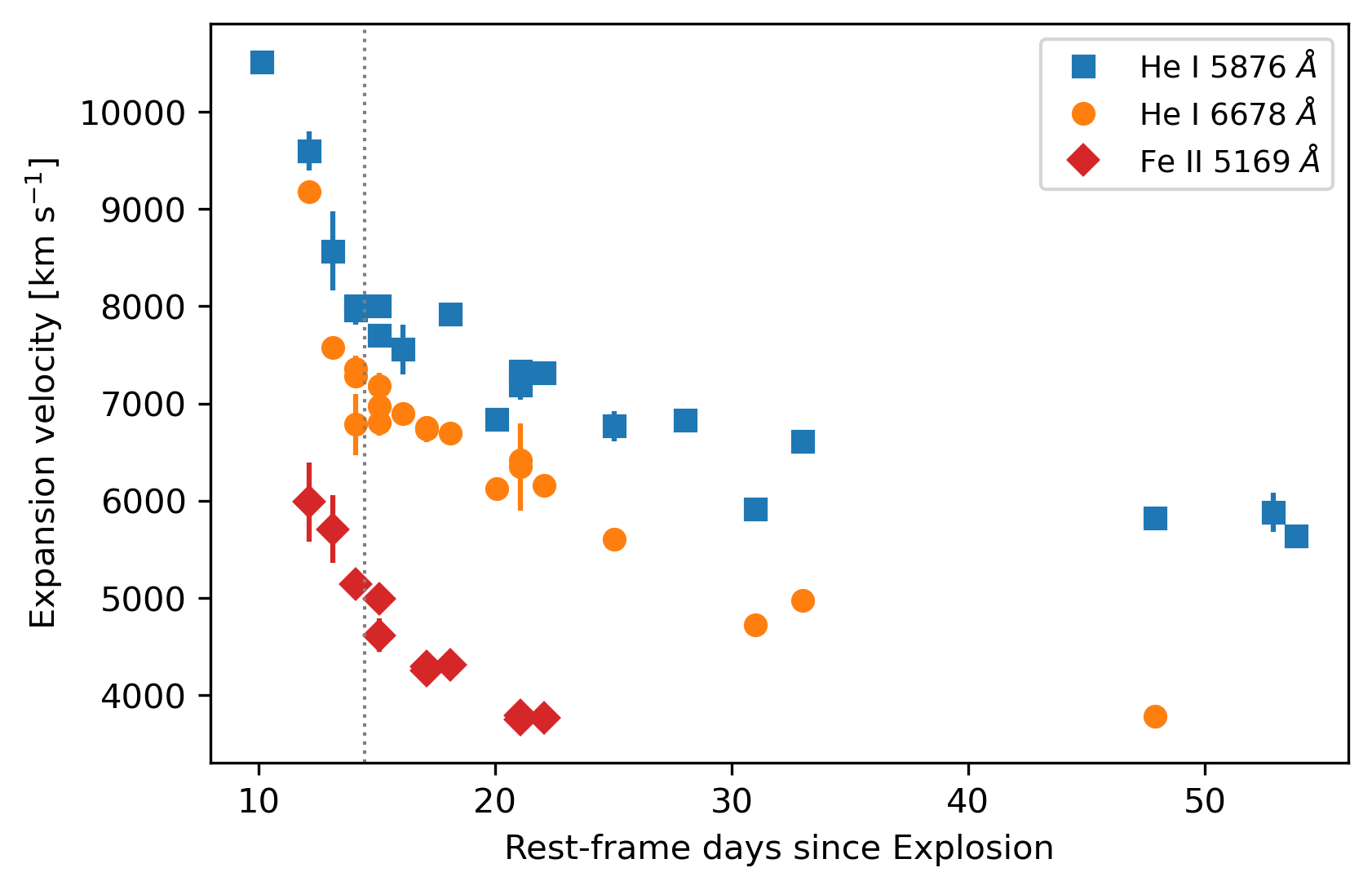}
    \caption{Line velocity evolution of SN~2021gno as derived from the minimum of the absorption components. For reference, the epoch of $r$-band maximum light is marked with dotted lines.}
    \label{fig:velo}
\end{figure}

We measured the expansion velocities of the prominent features in the spectra by fitting a Gaussian to the minimum of the absorption profiles. The velocity evolution of He~I~$\lambda$5876, He~I~$\lambda$6678, and Fe~II~$\lambda$5169 is shown in Figure~\ref{fig:velo}. For He~I~$\lambda$5876 the expansion velocity decreases from a value of $\approx$10,500 $\pm$ 120~km~s$^{-1}$ at 10 days after the explosion to $\approx$8000 $\pm$ 85~km~s$^{-1}$ at maximum light. After maximum the He~I~$\lambda$5876 velocity eventually levels off at approximately 6000 $\pm$ 90~km~s$^{-1}$ at 25 days from the explosion, whereas the He~I~$\lambda$6678 continues decreasing. For Fe~II~$\lambda$5169 the velocities go down from $\approx$5000 $\pm$ 370~km~s$^{-1}$ around maximum light to $\approx$3700 $\pm$ 90~km~s$^{-1}$ at 20 days after explosion, when it becomes too weak to be measured. 

\begin{figure}
	\includegraphics[width=\columnwidth]{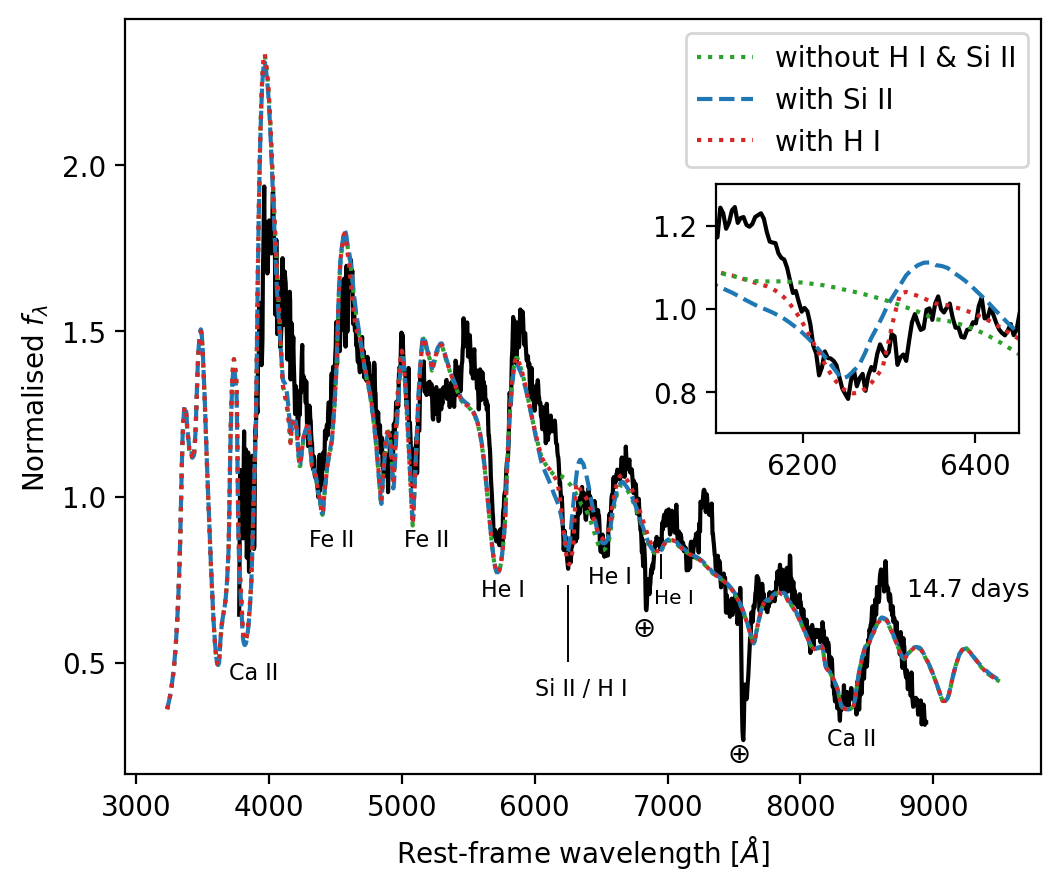}
    \caption{Comparison of the maximum-light spectrum of SN~2021gno at $14.7$ days after the explosion (solid black lines) with synthetic spectra calculated with SYNOW. Spectral features are labeled with the main ions that cause them. The SYNOW fit including Si~II and without hydrogen is plotted as a dashed blue line. An alternative calculation with hydrogen and without Si~II is shown with dotted red lines. For completeness, a model with no H$\alpha$ and no Si~II is shown (dotted green lines). The inset panel shows the detail around the observed feature at 6250~\AA, in which we indicated the Si~II~$\lambda$6355 line at photospheric velocity and the H$\alpha$ line detached at 15000~km~s$^{-1}$.}
    \label{fig:synow}
\end{figure}

In order to provide a robust identification of spectral features we calculated synthetic spectra using the SYNOW code \citep[]{fisher99,branch02} with the aim of reproducing the 14 days after explosion spectrum (i.e.\ around maximum light). The result is shown in Figure~\ref{fig:synow}. SYNOW assumes spherical symmetry, homologous expansion, a sharp photosphere at a given expansion velocity $v_{\mathrm{ph}}$\footnote{In this configuration, velocity is a measure of the distance of a given element of ejecta from the explosion center.} where the continuum emission is defined by a black body (BB) with temperature $T_{\mathrm{BB}}$, and it treats line formation with the Sobolev approximation. The observed spectrum was best reproduced with $T_{\mathrm{BB}}=7000$~K and $v_{\mathrm{ph}}=5000$~km~s$^{-1}$, and including He~I, O~I, Mg~II, Ca~II, Ti~II, and Fe~II in the ejecta (extra species are discussed below to explain the 6250~\AA\ feature). This list of species is not meant to be complete, as other elements or ions may produce identifiable lines. For each species an excitation temperature was adopted close to the assumed value of $T_{\mathrm{BB}}$, except for He~I and Ca~II, which required temperatures of about 15000 K in order to reproduce the relative line strengths. Such values should be considered only as a reference and not with strict physical meaning, as deviations from thermal equilibrium and other base assumptions are expected. 

The phostospheric velocity was suitable to reproduce the location of P-Cygni absorptions of most species by assuming a power-law distribution of optical depth as a function of velocity within the ejecta, with power-law indices ranging from 2 to 5 for different species. An exception to this was He~I, which was assumed to be distributed in a higher-velocity shell detached from the photosphere. This was done by adopting a Gaussian shape for the optical-depth distribution as a function of velocity coordinate, with center at 7500~km~s$^{-1}$ and width of $\sigma=3500$~km~s$^{-1}$. Additionally, in order to improve the match on the blue side of the Ca~II~H\&K and IR-triplet features, a high-velocity (HV) component was included for Ca~II, with a Gaussian distribution centered at 13000~km~s$^{-1}$ and with $\sigma=4000$~km~s$^{-1}$.   

With the aid of SYNOW we searched for species that could explain the absorption feature observed near 6250 \AA. We found a reasonable match by including Si~II with a similar distribution in velocity as the bulk of the elements. However, as shown in the inset of Figure~\ref{fig:synow}, the Si~II absorption at photospheric velocity appears slightly shifted to the blue of the observed line, although the discrepancy (of $\approx$500~km~s$^{-1}$) is not substantial enough to discard Si~II. Unfortunately, other Si~II transitions are blended and too weak to be unambiguously identified. This prevents a definitive conclusion to be driven as to the presence of Si~II. 

An alternative identification could be H$\alpha$ arising from a detached distribution of H~I with a central velocity of 15000~km~s$^{-1}$ and a width of $\sigma=2500$~km~s$^{-1}$ (also shown in the inset of Figure~\ref{fig:synow}). This is at a slightly higher velocity than that of the HV Ca~II component. The subsequent disappearance of the 6250~\AA\ feature would indicate that, if present, the amount and surface abundance of hydrogen in the outer part of the ejecta should be small \citep[H masses of $\sim$10$^{-3}-10^{-2}$~$M_\odot$; ][]{dessart11,hachinger12}. Again, it is not possible to unambiguously identify other H~I lines in the spectrum. Therefore the identification of this line remains uncertain, with the possibility of it being due to Si~II near the photospheric velocity, H$\alpha$ at high velocity, or a combination of both. In Section~\ref{sec:spec_neb} we examine the possible detection of H$\alpha$ emission in the nebular spectra of SN~2021gno, and in Section~\ref{sec:disc_concl} we further discuss the implications of this possibility for the progenitor scenario.

\begin{figure*}
	\includegraphics[scale=0.39]{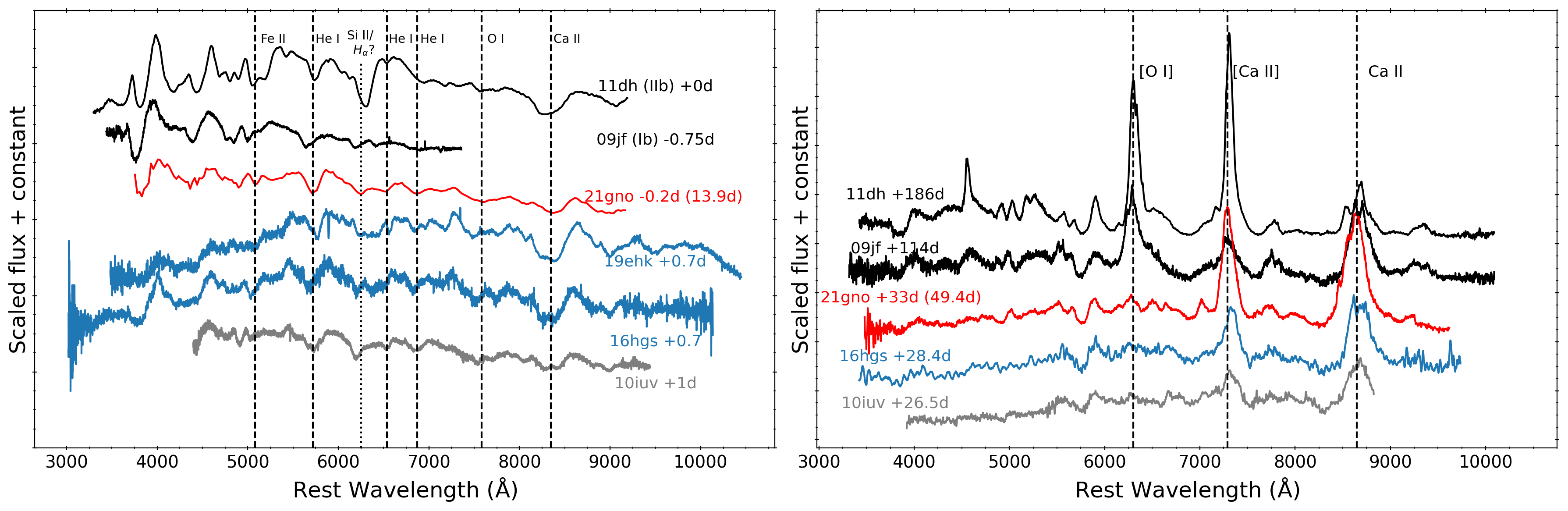}
    \caption{\textit{Left:} SN~2021gno spectral comparison at maximum light (red). The Ca-rich transient PTF10iuv \citep[]{kasliwal12} is marked in gray; the double peaked Ca-rich transients iPTF16hgs \citep[]{de2018}, and SN2019ehk \citep[]{de2021} are marked in blue. For comparison, the prototypical Type IIb SN2011dh \citep[]{ergon14} and Type Ib SN~2009jf \citep[]{modjaz14} are shown in black. Epochs are in rest-frame days since maximum light (for reference they are given relative to the explosion time in parentheses for SN~2021gno). Dashed vertical lines correspond to the main optical lines marked at the absorption minimum in the spectrum of SN~2021gno. The dotted vertical line is located at 6250~\AA~(see Section \ref{sec:spec_photospheric}). \textit{Right:} Spectral comparison of SN~2021gno at early nebular phase (in red). PTF10iuv is marked in gray, and the double peaked Ca-rich transient iPTF16hgs is marked in blue (references are the same as in the left panel). Type IIb 2011dh \citep[]{shivvers13} and Type Ib SN~2009jf \citep[]{shivvers19} spectra are marked in black.}
    \label{fig:spec_max}
\end{figure*}

In Figure~\ref{fig:spec_max} (left panel) the spectrum of SN~2021gno at maximum light is compared with those of other Ca-rich transients at a similar epoch, such as PTF10iuv, and the double-peaked events iPTF16hgs, and SN~2019ehk. As SN~2019ehk has substantial extinction from its host galaxy, we de-reddened its spectrum adopting a value of $E(B-V)_{\mathrm{Host}}=0.47$ mag from \citet[]{jacobsongalan20}, and using the parameterization of \citet[]{fitzpatrick99}. Given that Ca-rich transients are similar to type Ib/IIb SNe at this epoch, we have included the Type IIb SN~2011dh and Type Ib SN~2009jf at maximum light as they are representative within their respective classes and they have well-sampled spectra near maximum light. Note, however, that the epoch relative to the explosion in those cases is substantially larger due to their slower evolution. SN~2021gno looks similar to the other Ca-rich transients, particularly in the strength of the He~I features, which is not greatly different from standard Type Ib spectra. We note that there is an absorption commonly present at around 6200~\AA\ in Ca-rich transients that may be related with the $\approx$6250~\AA\ feature discussed above.

\subsubsection{Pre-nebular and nebular phase}
\label{sec:spec_neb}
In this section we study the spectra obtained after 30 days relative to the explosion time, when nebular transitions start to appear (see Figures~\ref{fig:spec} and \ref{fig:nebspec}). At these epochs the He~I features decrease in strength while Ca II and [Ca II] emissions become dominant. A weak [O~I]~$\mathrm{\lambda\lambda}$~6300,~6364 emission feature can also be seen in the last two spectra obtained after 110 days (see Figure~\ref{fig:nebspec}). He~I~$\lambda$5876 may still be detectable in the last two spectra, although it may be blended with the Na~I~D doublet. 

In Figure \ref{fig:spec_max} (right panel) we compare the spectrum of SN~2021gno entering the nebular phase with those of three other Ca-rich transients observed at similar epochs. SN~2021gno looks similar to the Ca-rich transients both in terms of the Ca~II and [Ca~II] emission and in the weak [O~I] emission. Except for the strength of the [O~I]~$\lambda\lambda$~6300,~6364 feature, the $\approx$50 d spectra of the Ca-rich objects resemble those of normal Type Ib and Type IIb SNe at a much later age of $\approx$$150-200$~d, i.e.\ during the nebular phase.

A weak emission is present at around 6550~{\AA} in the spectra obtained after 110 days (Figure~\ref{fig:nebspec}). Its identification is also ambiguous, as it could be produced by H$\alpha$ or by Ca~I]~$\mathrm{\lambda6572}$. [N~II]~$\lambda\lambda$6548,6583 may also contribute to this feature as proposed by \citet[]{jerkstrand15} \citep[see also][for a detailed analysis of this structure]{fang18}. We measured a central wavelength of $\approx$6540~\AA, which would correspond to velocity shifts of $\approx$$-1000$~km~s$^{-1}$, $\approx$$-1500$~km~s$^{-1}$ or $\approx$$-2000$~km~s$^{-1}$ if it were due to H$\alpha$, Ca~I] or [N~II] (relative to the strongest component at 6583~\AA), respectively\footnote{The OSIRIS spectrum in the observed configuration provides a resolution of $\approx$$300$~km~s$^{-1}$.}. The first shift is similar to what is seen for the [Ca II] $\lambda\lambda$~7291,~7324 feature (see below), so this favours the identification as H$\alpha$. On the other hand, the lack of noticeable H$\beta$ and the presence of strong [Ca~II] lines (Figure~\ref{fig:nebspec}) provide support to the association with Ca~I]~$\lambda$6572. In Section~\ref{sec:disc_concl} we further discuss the implications of the possible presence of hydrogen in the late-time spectra of some Ca-rich transients.

\begin{figure}
	\includegraphics[width=\columnwidth]{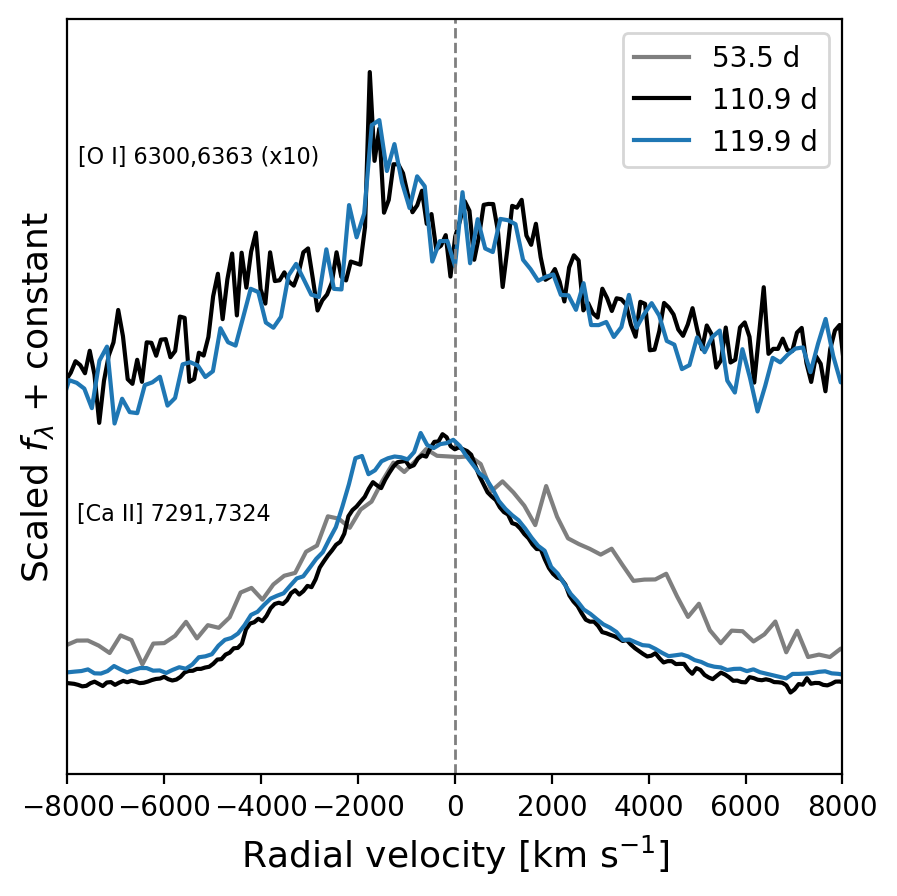}
    \caption{[Ca~II] and [O~I] emission lines observed in the transitional and nebular spectra of SN~2021gno, plotted in velocity space. The spectra correspond to $53.5$ days (gray), $110.9$ days (black), and $119.9$ days (blue) relative to the explosion epoch. Reference wavelengths were assumed as 6300~{\AA} for the [O~I]~$\lambda\lambda$6300,6363 complex, and 7304.4~{\AA} for the [Ca~II]~$\lambda\lambda$7291,7324 doublet. The [O~I] line of the $53.5$-days spectrum is not included since it is not visible at that epoch.}
    \label{fig:velprof}
\end{figure}

In Figure \ref{fig:velprof} we inspect the late-time profiles in velocity space of the [O~I]~$\lambda\lambda$6300,6364 and the [Ca~II]~$\lambda\lambda$7291,7324 emissions. The [O~I]~$\lambda\lambda$6300,6364 feature is compatible with a double-peaked shape \citep[]{mazzali2005}, which has been observed in nebular spectra of SE~SNe \citep[e.g.][]{prentice17}. A slight overall blueshift is seen with respect to the reference wavelength of 6300 \AA\footnote{The shift would be slightly larger if the $\lambda$6364 component influenced the actual reference wavelength of this blend.}. The blue peak is centered roughly at $-1300$~km~s$^{-1}$ while the red peak is centered at $\approx$800~km~s$^{-1}$. The separation between both peaks is substantially smaller than the one expected (of $\approx$3000~km~s$^{-1}$) between the line components. Therefore, the shape of this line is suggestive of an asymmetric distribution of the O-rich material, such as in a toroidal or a bipolar structure \citep[]{maeda08,tanaka09}. However the S/N is low, so this profile may be compatible with a shell-like structure result of the lack of oxygen in the innermost ejecta \citep[]{taubenberger09,mazzali2005,mazzali17}.

The [Ca~II] line is instead singly peaked, although it also appears to be blueshifted. Relative to the adopted effective central wavelength of 7304.4~\AA, the blueshift goes from $\approx$$-50$~km~s$^{-1}$ to $\approx$$-500$~km~s$^{-1}$ between 50 and 120 days past explosion. During the same time lapse, the line profile becomes asymmetric. As shown in Figure~\ref{fig:velprof}, the red side of the line at 111 and 120 days appears to be suppressed as compared with the 53-d spectrum. On the contrary, the shape of the line on the blue side remains nearly constant\footnote{Note that the wings of the line at 53 days are affected by the existence of non-negligible continuum emission.}. 

There are several possible explanations for the observed evolution of the [Ca~II] line. It could be due to changes in the timing of the emergence of contaminating lines, possibly from iron-peak elements \citep[]{maeda08,jerkstrand15,dessart21}. However, in order to maintain the blue side of the line nearly unchanged, such contaminants should grow and decrease on each side of the line in a highly coincidental fashion. Thus we consider this to be an unlikely possibility. Another explanation would be the formation of dust at some time between 50 and 110 days \citep[see, for example,][]{silverman13}. Dust would absorb the light from the far side of the ejecta and thus would reduce the emission on the red side of the line. The amount of dust would increase with time, causing an increasing blueshift of the line, which is observed \citep[][]{taubenberger09}. However, one would expect an increase of overall extinction and reddening at late times and therefore an increase of the light-curve decline slopes, especially in the bluest bands. This is not apparent in the observed light curves, at least until $\approx$80 days past explosion. Dust formation is thus unlikely the cause of the [Ca~II] line evolution. Nevertheless, some mechanism producing self absorption in a clumpy ejecta may reduce the flux on the red side of the line without a noticeable effect on the light curves and colours \citep[]{wang94}. 
Self absorption and scattering in the ejecta may also produce blueshifted, asymmetric lines, but the effect should decrease with time as the density decreases \citep[]{jerkstrand17}. The reduction in density is supported by the decrease in flux of  permitted features (e.g., the Ca~II IR triplet) relative to forbidden transitions ([Ca~II]~$\lambda\lambda$7291,7324), as can be seen in Figures~\ref{fig:spec_max} and \ref{fig:nebspec}. Thus the line should become more symmetric at later epochs, contrary to what we observe. A final alternative would be an asymmetric distribution of the Ca-rich material \citep[]{maeda07}, although this requires a change in the physical distribution of the material that produces the line occurring some time between 50 and 110 days. 

For the last two spectra, at 111 and 120 days after the explosion, we measured the [Ca~II]/[O~I] flux ratio, obtaining a mean value of $10.5 \pm 0.1$. This is a typical value for Ca-rich transients, as compared with those of normal Type Ib/c SNe that always lie below 2 \citep[]{milisavljevic17,fang22}. 

\section{Bolometric evolution and modeling}
\label{sec:bolmod}
To better understand the physics underlying SN~2021gno, we studied the photometric evolution via black body fits to the spectral energy distribution at each epoch. This allowed us to determine the BB temperature and radius evolution and to construct a bolometric light curve (Section~\ref{sec:bol}) that we then employed to compare with hydrodynamic models (Section~\ref{sec:mod}). 

\subsection{Bolometric luminosity}
\label{sec:bol}

After correcting the observed magnitudes for extinction (see Section~\ref{sec:dist_ext}), we converted them to monochromatic fluxes at the effective wavelength of each filter, using the transmission functions of the photometric system, taken from the Carnegie Supernova Project webpage\footnote{\url{https://csp.obs.carnegiescience.edu/data/filters}}. If a given epoch lacked observations in a certain band, we made interpolations using a Gaussian Processes method with the Python library GPy\footnote{\url{https://github.com/SheffieldML/GPy}}. In order to secure a self-consistent coverage in the optical range, we restricted the calculations to the photometry from the Swope telescope at early times, and the LCOGT at late times. We also interpolated the $z$-band, since it is necessary to obtain reliable BB fits. Then we integrated the monochromatic fluxes along wavelength for each epoch, using the trapezoidal method in order to obtain a  quasi-bolometric flux ($\mathrm{F_{qbol}}$). 
As $\mathrm{F_{qbol}}$ is calculated with all observed bands, and the wavelength coverage may vary between epochs, we also calculated the $\mathrm{F_{u \rightarrow z}}$ flux by consistently integrating in the range between the $u$ and $z$ bands. Because of the limited $u$-band coverage, $\mathrm{F_{u \rightarrow z}}$ was calculated between 2 and 24 days after the explosion. Both $\mathrm{F_{qbol}}$ and $\mathrm{F_{u \rightarrow z}}$ are listed in Table \ref{tab:lbol}.

To allow a comparison with the bolometric luminosities produced by our hydrodynamical models (Section~\ref{sec:mod}), we account for the flux outside the wavelength range covered by our broad-band photometry by performing ultraviolet and infrared extrapolations using BB fits to the spectral energy distribution at each epoch. At early epochs the BB distribution represents a good description of the SN emission, but as the ejecta expand and cool the emission at shorter wavelengths starts deviating from the BB model because of line blanketing produced by iron-group elements. Including the bluer bands in those cases worsens the BB fits and biases the temperatures toward lower values, which in turn produces an overestimation of the IR correction \citep[]{faran18,martinez22}. With the aim of having reliable fits, we followed a procedure similar to that of \citet[]{faran18}. When observations were available in either $w2$, $m2$, $w1$ or $u$ bands and the flux in those bands dropped by more than 1$\sigma$ below the BB model, we calculated it again excluding those bands. This was done while there were observations available in those bands ($\approx$16 days after the explosion in the UV, and $\approx$24 days in $u$ band). After that epoch, the fits included the complete set of optical bands.

We note that after around 30 days from the explosion, the presence of strong emission lines on the red part of the spectrum (see Figures~\ref{fig:spec} and \ref{fig:nebspec}) causes the SED to be overestimated in the $i$ and $z$ bands relative to the continuum. This, in turn, would produce an overestimation of the extrapolated IR flux. In order to estimate the size of this effect, we removed both strong calcium emission lines from the available spectra after 30 days past explosion, and we calculated synthetic photometry. The difference between the original and "continuum only" spectra at $\approx$30 days amounted to $0.1$ and $0.4$ mag in the $i$ and $z$ bands, respectively. At $\approx$50 days, the differences increased to $0.5$ and $0.95$ mag in the respective bands. To correct the observed $i$- and $z$-band photometry we used a straight-line fit to these differences between 30 and 50 days, and a constant value thereafter (as a conservative lower limit to the correction given the lack of spectra until 110 days). This was done solely to produce corrected SEDs to be fit by BB functions for extrapolation, as the integrated fluxes $\mathrm{F_{qbol}}$ and $\mathrm{F_{u \rightarrow z}}$ were left unmodified.

The infrared flux beyond the $z$ band, $\mathrm{F_{IR}}$, was estimated by integrating the fitted BB function from the effective wavelength of the $z$ band to infinity. To account for the unobserved ultraviolet flux, $\mathrm{F_{UV}}$, when all bands were available for the BB fits, we integrated the BB from 0 {\AA} to the bluest band observed at that epoch. In the rest of the cases we integrated a linear extrapolation from the effective wavelength of the bluest band to zero flux at 2000 {\AA} \citep[]{bersten09}.

\begin{figure}
	\includegraphics[width=\columnwidth]{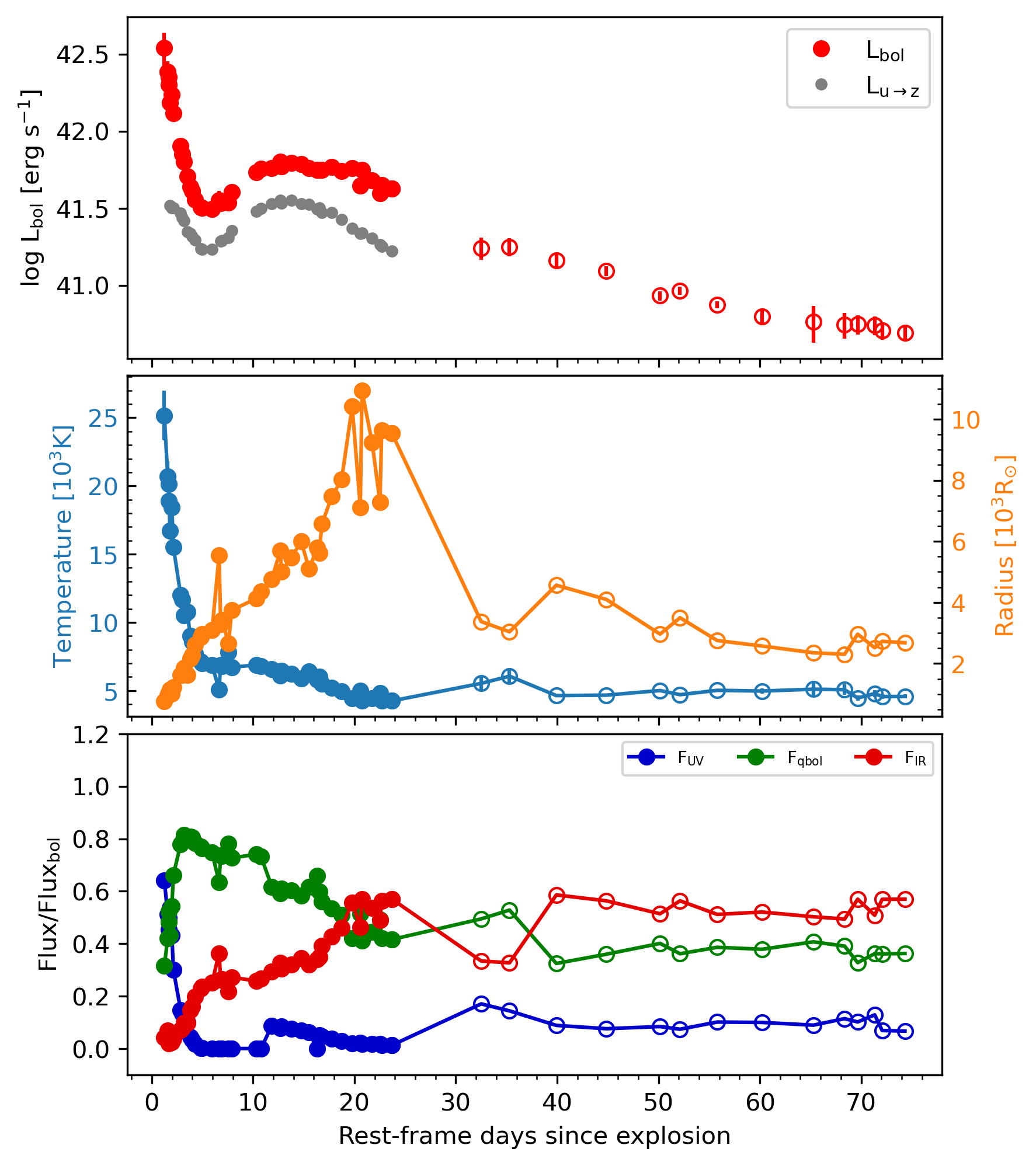}
    \caption{\textit{Top:} Bolometric light curve of SN~2021gno. The luminosity calculated from $\mathrm{F_{u \rightarrow z}}$ is also shown. \textit{Middle:} Evolution of the fitted BB temperature (blue) and radius (orange) for SN~2021gno. \textit{Bottom:} Contributions to the total flux. $\mathrm{F_{qbol}}$ is the integrated flux in the available bands, while $\mathrm{F_{UV}}$ and $\mathrm{F_{IR}}$ are the estimated flux from extrapolations to the UV and IR, respectively (see text for details). Open symbols indicate epochs where $z$-band was not available. }
    \label{fig:trev}
\end{figure}

The total bolometric fluxes were calculated as $\mathrm{F_{bol}}=$ $\mathrm{F_{UV}} + \mathrm{F_{qbol}} + \mathrm{F_{IR}}$ and then converted to luminosity assuming the distance calculated in Section~\ref{sec:dist_ext}. The uncertainty in the luminosity was estimated by considering uncertainties in the distance, photometry, and the estimated errors of the extrapolated fluxes. The resulting bolometric light curve is shown on the top panel of Figure~\ref{fig:trev}. The luminosity drops rapidly during the first $\approx$5 days past explosion before rising to a local maximum that occurs at $\approx$15 days. Due to the lack of $z$-band data and the uncertainties in the IR extrapolations after about 25 days past explosion, we consider the bolometric light curve to be less reliable after that epoch than at earlier times.

The BB temperature and radius evolution are shown on the middle panel of Figure~\ref{fig:trev}. At early epochs, the BB temperature drops rapidly, from over 20000~K to $\approx$7000~K in less than five days, which is indicative of a fast cooling of the shocked envelope. This fast cooling coincides with the initial light-curve peaks. We note the significance of UV observations to make reliable estimations of the temperature at early times. After that, the temperature shows a slower decline, reaching $\approx$4000~K at times past maximum light. Both behaviours are consistent with the colour evolution of SN~2021gno seen in Figure~\ref{fig:ex_comp}. At later times (45 to 75 days past explosion), the temperature remains nearly constant, although we note that the BB approximation breaks down as the SN transitions into an emission-line dominated spectrum. 

Similar to the temperature evolution, the BB radius shows a steep initial rise until about day 5 (with a velocity of $\approx$5000~km~s$^{-1}$), and then slows down to $\approx$3000~km~s$^{-1}$ until about $\approx$25 days after the explosion. Until that time, despite the aforementioned blanketing effects, the BB approximation remains valid and the BB radius roughly follows the photospheric radius. Later on the derived BB radius recedes to a nearly constant value, although this behaviour probably lacks any physical meaning. Although the methods employed in \citet[]{jacobsongalan22} to compute the bolometric luminosity are slightly different, all the estimated bolometric quantities are similar to our results. 

The fractional contributions of the UV/$\mathrm{F_{qbol}}$/IR components to the total flux are depicted on the bottom panel of Figure~\ref{fig:trev}. The observed contribution ($\mathrm{F_{qbol}}$) dominates between about three and 15 days past explosion (i.e.\ from slightly before the bolometric local minimum until slightly past maximum light). At earlier epochs, when the ejecta are very hot, the extrapolated UV flux is significant and it contributes up to 60\% of the total estimated flux. The UV contribution rapidly decreases as the temperature drops. The contribution of the IR extrapolated flux is small at early times but it becomes important soon after and it dominates after maximum light. We estimate it to be $\approx$60\% of the total flux at the latest epochs, although as mentioned above, the IR flux may be overestimated. 

\subsection{Hydrodynamical modeling}
\label{sec:mod}
We model the bolometric light curve and the line velocity evolution of SN~2021gno with the 1-dimensional Lagrangian hydrodynamical code of \citet[]{bersten11}. In the context of a calcium-rich transient, we propose an ad-hoc pre-explosion model with a highly stripped star as an initial condition. In order to obtain the initial luminosity peak, the additional presence of circumstellar matter (CSM) is assumed. Even though the model simulation is self-consistent at all epochs, the analysis can be decoupled into two distinct phases: a) the initial decline due to post-shock cooling of the CSM, and b) subsequent evolution through the main peak that is powered by radioactive decay. From the former phase we can mainly obtain the mass and extent of the CSM, and from the latter phase one can derive other overall explosion parameters, such as the ejected mass, the explosion energy and the mass of radioactive material.

As a pre-explosion structure for the main peak of the light curve we construct a highly stripped object by using the public stellar evolution code MESA\footnote{\url{http://mesa.sourceforge.net/}} version 15140 \citep[]{paxton11, paxton13, paxton15, paxton18, paxton19} with the `input files' of \citet[]{moriya17} modified to comply with the new MESA version and to include a nuclear network of 21 isotopes (`approx21.net'). The prescription proposed by \citet[]{moriya17} simulates mass loss from a He star via Roche-lobe overflow (RLOF) in a simplified manner, following the evolutionary calculations by \citet[]{tauris13} of a close binary system comprising a He star that transfers matter to a neutron star.

According to \citet{tauris13}, mass transfer by RLOF occurs after core He exhaustion. Therefore, we first evolve an isolated He star of 4 $M_{\odot}$ until it finishes burning He in its core. Then, we reduce the mass of the star using the parameter `relax mass', defined by MESA, until the He star reaches a mass of $2.5$~$M_\odot$ and we continue the evolution until core collapse, which is taken as the time when any location inside the stellar model reaches an infall velocity of 1000~km~s$^{-1}$. At the end of its evolution, the star has $2.48$~$M_\odot$ and a CO core mass of $1.89$~$M_\odot$. This core mass would inidicate a $M_{ZAMS} \approx 15-17 M_{\odot}$ \citep[]{ertl20}, although this may not apply to the atypical evolutionary path considered here. To this compact structure we attach different forms of He-rich CSM with wind-like density distributions of varying mass and radial extent. 

\begin{figure*}
	\includegraphics[scale=0.80]{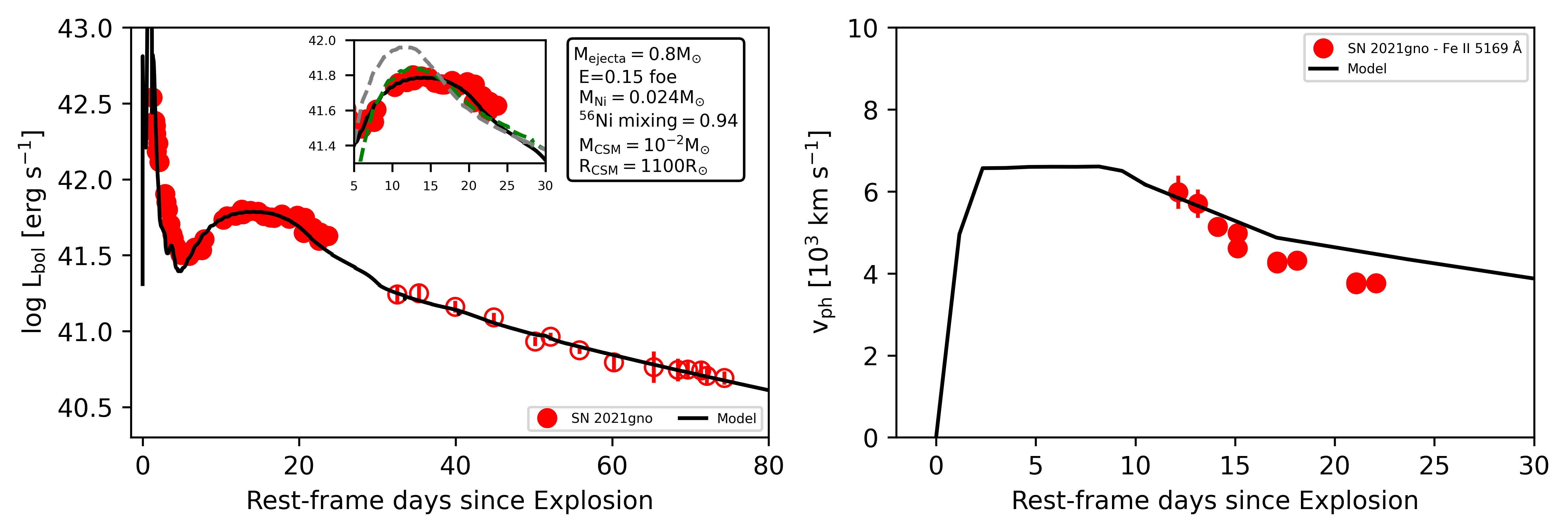}
    \caption{The bolometric light curve of SN~2021gno (red dots, left panel) and Fe II velocity evolution (red dots, right pannel), compared with the hydrodynamical model discussed in Section \ref{sec:mod} (solid black lines). Open symbols indicate epochs where $z$-band was not available. The parameters of the model and the CSM parameters that reproduce the early data are shown in the box. $\mathrm{^{56}Ni\;mixing}$ is given in fraction of the pre-SN mass. The inset plot in the left panel shows two alternative models (dashed lines) around maximum light for a lower-mass pre-SN progenitor of $2.2$~$M_{\odot}$, with the same conditions as the preferred model (gray), and varying the energy and $^{56}$Ni mass (green).}
    \label{fig:model}
\end{figure*}

To initialize the explosion, some energy is deposited in the form of a thermal bomb at a given mass `cut' within the pre-SN structure. In this case the mass cut was set to $1.7$ $M_\odot$, a value that is somewhat higher than the composition interface between the silicon core and oxygen core (see \citealp[]{morozova18}), and it was assumed to form a compact remnant. The ejected mass was therefore of $0.8$~$M_\odot$. We varied the explosion energy and $^{56}$Ni mass to reproduce the main light-curve peak and the evolution of the Fe II velocity. The resulting preferred model, shown in Figure~\ref{fig:model}, has an explosion energy of $E\,=\,0.15\,\times\,10^{51}$~erg and $^{56}$Ni mass of $M_\mathrm{^{56}Ni}\,=\,0.024~M_{\odot}$. The degree of Ni mixing was fixed to an extension of 95\% of the progenitor mass. This parameter affects the timing and depth of the light-curve minimum as more extended nickel produces an earlier rise to the main maximum \citep[]{bersten12}. In order to match the bolometric luminosity after day 30 we needed to lower the model luminosity by artificially reducing the gamma-ray opacity from the standard value of $\kappa_{\gamma} = 0.03$~cm$^2$~g$^{-1}$ to $\kappa_{\gamma} = 0.01$~cm$^2$~g$^{-1}$. Such an ad-hoc modification of the gamma opacity has been used in the literature \citep[]{tominaga05,folatelli06,gutierrez21,gutierrez22} to mimic a reduction of the gamma-ray deposition in the ejecta when the nickel mass required to explain the peak luminosity overestimates the tail luminosity. Nevertheless, we note that the late-time bolometric luminosity is less reliable than around maximum light (see Section~\ref{sec:bol}). In any case, this modification does not affect the conclusions of our analysis.
  
We note that the ejected mass derived from the main peak of the light curve is somewhat larger than typical masses assumed for ultra-stripped progenitors, which are $M_\mathrm{ej}<0.2$~$M_\odot$ \citep[]{tauris15}. However, when we adopted lower-mass pre-SN models (all starting from the same 4 $M_{\odot}$ He star), the resulting light curves were too narrow compared with the observations around the main maximum, as shown in the inset plot of Figure \ref{fig:model}.

Based on the overall explosion parameters found above we focused on fitting the early behaviour as arising from the cooling of the shocked CSM. For this purpose, we modified the external density structure of the progenitor before injecting the explosion energy. We assumed density profiles of the type $\rho \sim r^{-2}$ for the CSM, i.e. due to a steady wind, with different extensions and masses.
In Figure~\ref{fig:model}, we present a model with a CSM mass of $\sim\,10^{-2}\,M_{\odot}$ and a radius of $1100\,R_\odot$ that reproduces very well the observations. In the case of a steady wind, this model corresponds to a mass-loss rate of $4\,\times\,10^{-3}\,M_{\odot}$~yr$^{-1}$. We note that a similar cooling behaviour may be obtained by assuming different density distributions of the CSM. For instance, an accelerated wind \citep[][]{moriya17} or an extended envelope in hydrostatic equilibrium were not tested in this case. 

The model presented here is not necessarily the only possible scenario for this SN. However, the calculations indicate that this setup of a highly stripped star with a tenuous material around it is plausible. The parameters quoted here may serve as a comparison for similar objects. The nature of the surrounding material, whether a wind or the result of binary mass transfer or pre-explosion eruptions, is not clear. Nevertheless, the initial decline of the bolometric luminosity and the fast drop of the black-body temperature (Figure~\ref{fig:trev}) strongly suggest the presence of such a material.

\section{Physical origin of double-peaked Ca-rich transients}
\label{sec:disc_concl}

 Along with SN~2021gno, other Ca-rich objects that showed double-peaked light curves are iPTF16hgs \citep[]{de2018}, SN~2018lqo \citep[]{de2020}, SN~2019ehk \citep[]{jacobsongalan20,de2021,nakaoka21}, and SN~2020inl \citep[]{jacobsongalan22}. In Table~\ref{tab:double-peaked} we summarize the environmental properties of these events. We have included in the table the related transient iPTF14gqr \citep[]{de2018b}; although it is not generally considered as a Ca-rich event. It should be noted that some Ca-rich transients lack early-time observations, and they may have a first peak that was not observed. \citet[]{jacobsongalan22} point out that 5 out of 9 (55\%) Ca-rich transients discovered less than 3 days after explosion and with $<$2 day cadence in the observations show this early peak in the light curve. However, they also open the possibility of marginal detections of very early first peaks for the rest of the sample, meaning that this feature could be present in all Ca-rich transients.
 
 With the exception of SN~2018lqo and SN~2021inl, the rest of the double-peaked Ca-rich transients appeared in environments where ongoing star formation is plausible. This opens the possibility that a subgroup of Ca-rich transients with double-peaked light curves arises from a younger population than that of WD stars, i.e., this group may originate from core-collapse  explosions of massive stars. \citet[]{de2021} suggested that they can arise from stripped stars at the low-mass end of CCSN progenitors ($8-10\,M_\odot$). \citet[]{nakaoka21} also proposed a core-collapse origin for iPTF14gqr, iPTF16hgs, and SN~2019ehk, within the more extreme USSN scenario. For SN~2021gno we presented a progenitor setup consistent with a highly-stripped massive star with an ejecta mass of $0.8\,M_\odot$ and a $^{56}$Ni mass of $M_\mathrm{^{56}Ni}\,=\,0.024~M_{\odot}$. Although the ejecta mass is higher than what was estimated for those other objects, the proposed scenario is similar. In conclusion, both WD and massive-star progenitor scenarios may coexist in the case of double-peaked Ca-rich transients.
 
 In the core-collapse scenario the presence of the initial light-curve peaks is explained by the existence of an extended CSM or a thin extended stellar envelope. Furthermore, the detection of X-ray emission in the early light curve of both SN~2019ehk \citep[]{jacobsongalan20} and SN~2021gno \citep[]{jacobsongalan22}, is consistent with the presence of a shocked CSM. In the case of SN~2021gno, we proposed a CSM containing $\sim$$10^{-2}\,M_{\odot}$ and an extension of $1100\,R_{\odot}$. 
 
 Double-peaked light curves may also be found in the WD scenario as a consequence of He-shell detonations \citep[][]{shen10}. In fact, by modelling X-ray observations of SN~2021gno, and using shock cooling, shock interaction, and radio models, \citet[]{jacobsongalan22} concluded that the progenitor CSM density is consistent with models for the merger of low-mass, hybrid WDs. They further favour a WD origin for SN~2021gno, contrary to a core-collapse scenario, based on a series of arguments that we address next.

Firstly, \citet[]{jacobsongalan22} state that ultra-stripped stars produce too low ejecta masses and too little He to explain the observations of SN~2021gno. However, a less extreme progenitor such as the one presented here does produce the necessary ejecta mass (tenths of $M_\odot$) to comply with the light curve and enough He ($\approx\,0.5\,M_\odot$) to match the observed Type-Ib spectrum \citep[see][]{hachinger12}. \citet[]{jacobsongalan22} point out that all of the binary evolution models in the comprehensive set of BPASS \citep[]{eldridge17} calculations would produce too large ejecta masses for SN~2021gno. Our calculations, however, prove the viability of a low-enough ejecta mass as a consequence of an atypical binary evolution path similar to that of USSN progenitors \citep[]{tauris13,moriya16}.

Another point raised by \citet[]{jacobsongalan22} is the lack of radio detections, which would be inconsistent with a high mass-loss rate necessary to remove the H-rich envelope of a massive progenitor. We note, however, that in our proposed scenario, the H-rich material is lost upon leaving the main sequence. That is, at least $\sim\,10^4$ years before the explosion. At an assumed expansion velocity of 100~km~$s^{-1}$, this material would lie beyond $\sim\,10^{18}$~cm from the progenitor, i.e.\ far beyond the distance of $\sim\,10^{16}$~cm that is probed by the radio constraints.
 
Lastly, \citet[]{jacobsongalan22} favour a WD over a core-collapse origin for SN~2021gno also based on the low star-formation rate derived from a limit on the local H$\alpha$ luminosity of $L_{\mathrm{H}\alpha}\,<\,4.3\,\times\,10^{36}$~erg~s$^{-1}$. However, such a limit is still compatible with what is observed for a substantial fraction of normal SESNe, as shown by \citet[]{kuncarayakti18}. Therefore a massive-star origin for SN~2021gno cannot be ruled out based on this star-formation rate constraint.

\begin{table*}
\caption{Properties of double-peaked Calcium-rich transients.}
\label{tab:double-peaked}
\begin{threeparttable}
	\begin{tabular}{ccc|p{45mm}} 
		\hline
		SN & Host type/ Projected offset from the host nucleus & Star-forming region & Proposed progenitor \\
		\hline
		iPTF14gqr & Interacting double system / 29 kpc & Maybe & -USSN \citep[]{de2018b}\\
		iPTF16hgs & Dwarf spiral galaxy\tnote{a} / 5.9 kpc & Yes & -CCSN from highly stripped massive star in a close binary system \citep[]{de2018} \newline -He-shell detonations on WDs \citep[]{de2018}\\
		SN 2018lqo & E-type galaxy / 15.46 kpc & No & - \\
		SN 2019ehk & SAB(s)bc / 1.8 kpc & low SFR & -Merger of low-mass WDs \citep[]{jacobsongalan20} \newline -USSN from He star + NS system \citep[]{nakaoka21} \newline -CCSN from low-mas stripped progenitor \citep[]{de2021} \\
		SN 2021gno & SBa galaxy / 4.5 kpc & low SFR & -Merger of low-mass, hybrid WDs \citep[]{jacobsongalan22} \newline -CCSN from stripped progenitor (this work) \\
		SN 2021inl & E/S0 galaxy / 23.3 kpc  & No & Merger of low-mass, hybrid WDs \citep[]{jacobsongalan22}\\
    	\hline
	\end{tabular}
	\begin{tablenotes}
    \item[a] The galaxy name is not reported, yet a detailed analysis of its properties can be found in \citet[]{de2018b}.
    \end{tablenotes}
    \end{threeparttable}
\end{table*}

Notwithstanding the above considerations, we note that the massive star progenitor model for SN~2021gno we propose presents some caveats to be taken into account. Firstly, a more detailed binary evolution model should be performed to confirm that a low-mass, low-Ni SN is indeed possible. 
Second, our model has an initial He-core mass that would imply a relatively massive progenitor, which could conflict with the lack of H$\alpha$ emission and thus low SFR at the SN location. This limit is compatible with a low though non-negligible fraction ($\approx\,20\%$) of the stripped-envelope SN population.

\begin{figure}
	\includegraphics[width=\columnwidth]{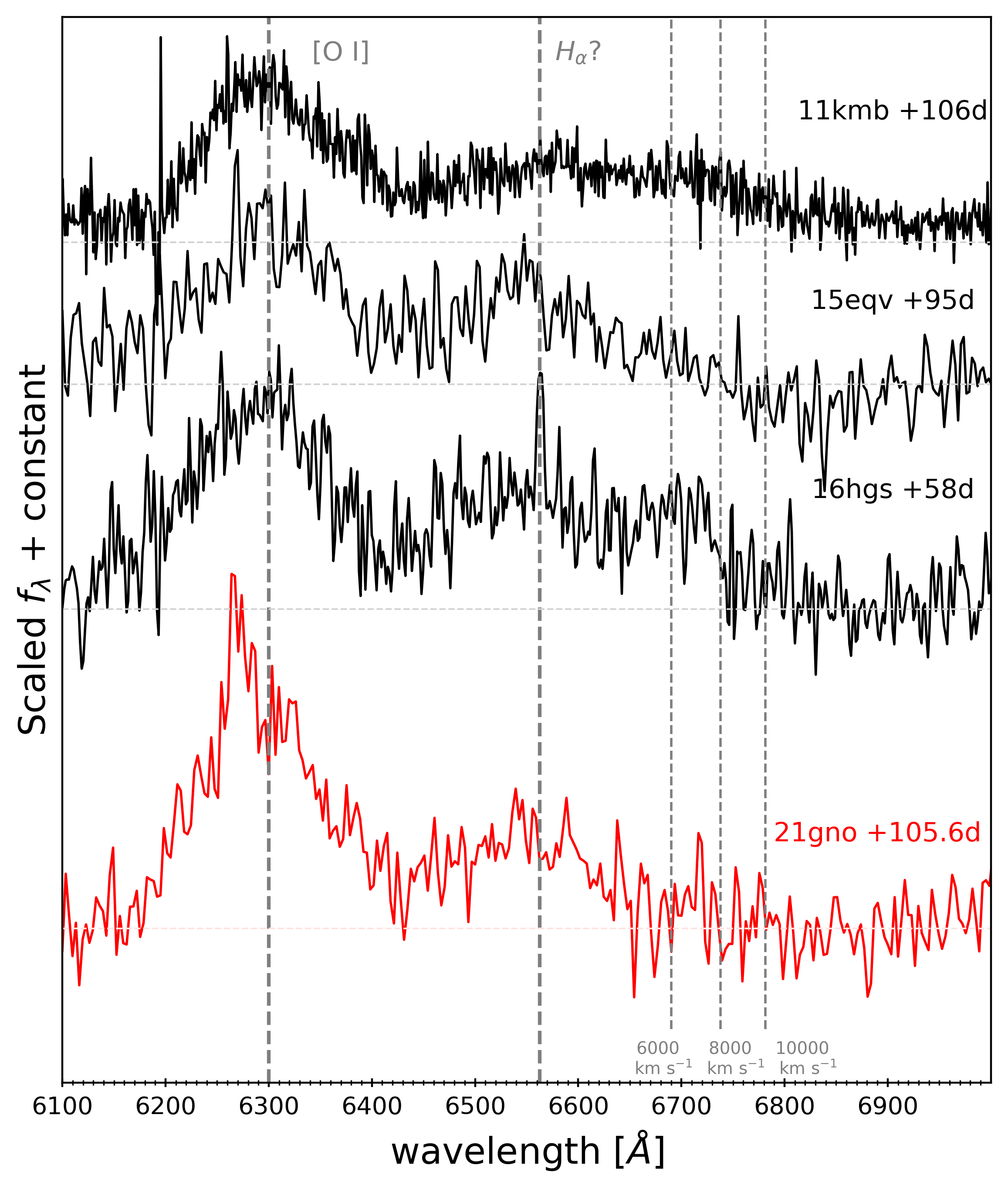}
    \caption{Comparison of the nebular spectrum of SN~2021gno (in red) with those of other Ca-rich transients (in black) around the [O~I]~$\lambda\lambda$6300,6363 / H$\alpha$ region. Transients PTF11kmb \citep[]{lunnan17}, iPTF15eqv \citep[]{milisavljevic17}, and iPTF16hgs \citep[]{de2018} are shown with their epochs relative to maximum labeled. Dashed horizontal lines indicate the continuum level in the range of $6850 - 7000$~\AA. Dashed vertical lines indicate the reference wavelengths of 6300~\AA\ for the [O~I] doublet, and $6562.8$~\AA\ for H$\alpha$. Dotted vertical lines show the position of H$\alpha$ at 6000, 8000, and 10,000~km~s$^{-1}$.}
    \label{fig:neb_halpha}
\end{figure}

We have also investigated the possible presence of hydrogen features in the spectra of SN~2021gno. Our SYNOW analysis suggests the possible presence of H$\alpha$ near maximum light, although the identification is not certain. \citet[]{de2021} claim the detection of H features for SN~2019ehk at very early phases (from flash spectroscopy), near maximum light, and at nebular phases.  Interestingly, the nebular phase spectra of SN~2021gno do show a bump that is centered at the location of H$\alpha$, as shown in Figure~\ref{fig:neb_halpha}. The figure shows that similar features are common among double-peaked Ca-rich transients. If they were produced by H$\alpha$, these emissions would typically extend out to approximately $8000 - 10000$~km~s$^{-1}$. In the case of SN~2021gno, the line extends roughly between 6000 and 8000~km~s$^{-1}$, in good agreement with the extent of the [Ca~II] and [O~I] lines (see Figure~\ref{fig:velprof}). However, a definitive identification for this weak emission is not possible, as it may be due to Ca~I]$\,\lambda$6572, to [N II]$\,\lambda\lambda$6548,6583, or to a combination of both \citep[]{milisavljevic17}. 

The presence of hydrogen in Ca-rich transients is a key point since it is directly related with the progenitor system and its ability to retain some H-rich material. If confirmed, such noticeable features in the maximum-light spectra would imply hydrogen masses of $\sim$10$^{-3}-10^{-2}$~$M_\odot$ \citep[][]{dessart11,hachinger12}.
The proposed scenarios involving the explosion of a WD star would be difficult to reconcile with this observation, as the hydrogen mass should be below $10^{-4}$~$M_\odot$ \citep[]{zenati19}. A more natural explanation may be that of a relatively massive progenitor leading to the collapse of a stripped core surrounded by an extended envelope or CSM containing traces of hydrogen. If the observed features are due to H$\alpha$, then SN~2021gno would belong to a small Type IIb class of Ca-rich transients. Confirmation of the presence of H requires synthetic spectra calculations applied to the explosion scenarios proposed for this type of transients, and specifically in the low ejected mass regime of $M_\mathrm{ej}\lesssim1\,M_\odot$. 

It is worth noting that this discussion is valid for the sub-sample of double-peaked Ca-rich transients we are considering here. Nevertheless, for other Ca-rich events, such as SN~2018lqo and SN~2021inl, WD progenitors are favored and a massive star origin would not be suitable. As we discussed above, in the case of SN~2021gno, it appears that both formation channels are plausible. While the co-existence of two distinct progenitor channels may introduce additional complexity to the overall scenario, it is worth contemplating both until further observational and modeling efforts are done to shed more light on the matter.

\section{Conclusions}
We have presented follow-up observations of SN~2021gno obtained by the POISE project and collaborators. According to its photometric and spectroscopic properties, this event belongs to the calcium-rich transient class. This is determined by its low luminosity, rapid photometric evolution, early evolution to the nebular phase and strong Ca emission in its nebular spectra. The very early UV and optical data showed double-peaked optical light curves, and extremely fast initial declines in the UV bands. The presence of an initial light-curve maximum is a characteristic that is shared with only a small number of calcium-rich transients. In addition, SN~2021gno appeared well within its host galaxy (in projected location), as opposed to the bulk of the calcium-rich class. 

At maximum light, the spectroscopic properties of SN~2021gno are similar to those of Type Ib SNe, as most Ca-rich transients. We calculated synthetic spectra using SYNOW to identify spectral lines. Particularly, we analyzed the feature near 6250 \AA~ and discussed the possibility of it being due to H$\alpha$ at high velocity, Si~II at photospheric velocity, or a combination of both. Additionally, we analyzed the nebular spectra, finding a double-peaked [O~I]~$\lambda\lambda$6300,6364 profile and a blue-shifted [Ca~II]~$\lambda\lambda$7291,7324 profile. There is a weak emission around 6250 \AA~ in the nebular spectra, whose identification is also ambiguous. It could be produced by H$\alpha$, Ca~I], or [N~II]. We hope future observations of Ca-rich transients will explore the H identification since its presence or lack thereof is crucial to understand the progenitor configuration.

We present a hydrodynamical model that reproduces the bolometric light curve and line velocity evolution. We found that the observations are compatible with the explosion of a highly-stripped massive star with an ejecta mass of $0.8\,M_\odot$ and a $^{56}$Ni mass of $0.024~M_{\odot}$. The initial cooling phase is explained by the presence of an extended CSM containing $\sim$$10^{-2}\,M_{\odot}$ with an extension of $1100\,R_{\odot}$. This opens the possibility of two different progenitor channels that may coexist to explain the origin of double-peaked Ca-rich transients.

SN~2021gno is yet another example that indicates that high-cadence early observations, as well as deep late-time observations, are crucial for determining the physical origin of Ca-rich transients and the external properties of their progenitors. Continuing such rapid follow-up efforts will help to understand whether or not early peaks in the light curves are intrinsic to this type of transients.

\section*{Acknowledgements}
This work was funded by ANID, Millennium Science Initiative, ICN12$\_$009. L.M. acknowledges support from UNRN~PI2022~40B1039 grant. M.O. acknowledges support from UNRN~PI2022~40B1039 and grant PICT-2020-SERIEA-01141. E.B. and J.D. are supported in part by NASA grant 80NSSC20K0538. 
L.G. acknowledges financial support from the European Social Fund (ESF) "Investing in your future" under the 2019 Ram\'on y Cajal program RYC2019-027683-I. L.G. and T.E.M.B. acknowledges financial support from the Spanish Ministerio de Ciencia e Innovaci\'on (MCIN), the Agencia Estatal de Investigaci\'on (AEI) 10.13039/501100011033 under the PID2020-115253GA-I00 HOSTFLOWS project, and from Centro Superior de Investigaciones Cient\'ificas (CSIC) under the PIE project 20215AT016. L.G., T.E.M.B. and N.E.R. acknowledges partial support from the program Unidad de Excelencia María de Maeztu CEX2020-001058-M. B.J.S. is supported by NSF grants AST-1920392, AST-1911074, and AST-1911074. The Aarhus supernova group is funded by the Independent Research Fund Denmark (IRFD, grant number 8021-00170B and 10.46540/2032-00022B), and by the VILLUM FONDEN (grant number 28021). N.E.R. acknowledges partial support from MIUR, PRIN 2017 (grant 20179ZF5KS), from the Spanish MICINN grant PID2019-108709GB-I00 and FEDER funds. H.K. was funded by the Academy of Finland projects 324504 and 328898. T.E.M.B. acknowledges financial support from Centro Superior de Investigaciones Cient\'ificas (CSIC) under the I-LINK 2021 LINKA20409. J.T.H. is supported by NASA award 80NSSC22K0127. L.K. is supported by the Hungarian National Research, Development and Innovation Office grant PD-134784. T.S., L.K., and K.V. are supported by the J\'anos Bolyai Research Scholarship of the Hungarian Academy of Sciences. R.K.T., T.S., and K.V. are supported by the \'UNKP 22-4, and Bolyai+ grant 22-5 and \'UNKP-22-5-ELTE-1093 New National Excellence Programs of the Ministry for Culture and Innovation from the source of the National Research, Development and Innovation Fund, respectively. M.G. is supported by the EU Horizon 2020 research and innovation programme under grant agreement No 101004719. M.N. is supported by the European Research Council (ERC) under the European Union’s Horizon 2020 research and innovation programme (grant agreement No.~948381) and by a Fellowship from the Alan Turing Institute.

The POISE collaboration is grateful to the Carnegie TAC for generous time allocation on the Swope telescope and to Las Campanas technical staff for excellent assistance during the observations. Based on observations collected at the European Southern Observatory under ESO programmes 1103.D-0328 and 0105.D-0511, and on observations made with the Gran Telescopio Canarias (GTC), installed in the Spanish Observatorio del Roque de los Muchachos of the Instituto de Astrofísica de Canarias, in the island of La Palma. 
This project has received funding from the European Union’s Horizon 2020 research and innovation programme under grant agreement No 101004719. This project has been supported by the GINOP-2-3-2-15-2016-00033 project of the National Research, Development and Innovation Office of Hungary (NKFIH) funded by the European Union, as well as by NKFIH grants OTKA FK-134432, 2019- 2.1.11-T\'ET-2019-00056, OTKA K131508 and the \'Elvonal grant KKP-143986. Authors acknowledge the financial support of the Austrian-Hungarian Action Foundation (101\"ou13).

%%%%%%%%%%%%%%%%%%%%%%%%%%%%%%%%%%%%%%%%%%%%%%%%%%
\section*{Data Availability}

The photometry presented in this paper is available in \cref{tab:phot_opt,tab:phot_opt_bajakonk,tab:phot_opt_lcogt,tab:phot_uv} and the spectra via the WISeREP\footnote{\url{https://www.wiserep.org/}} archive \citep[]{yaron12}.


\bibliographystyle{mnras}
\bibliography{2021gno} % if your bibtex file is called example.bib

%%%%%%%%%%%%%%%%%%%%%%%%%%%%%%%%%%%%%%%%%%%%%%%%%%

%%%%%%%%%%%%%%%%% APPENDICES %%%%%%%%%%%%%%%%%%%%%

\appendix
\section{Tables}
\begin{table*}
	\caption{Optical photometry of SN 2021gno with Swope Telescope}
	\label{tab:phot_opt}
	\begin{tabular}{lccccccc} 
		\hline
		Date & JD & $u$ & $B$ & $g$ & $V$ & $r$ & $i$ \\
		\hline
		2021 Mar 21 & 2459294.60 & $17.35 \pm 0.03$ & $17.53 \pm 0.01$ & $17.52 \pm 0.01$ & $17.62 \pm 0.02$ & $17.74 \pm 0.01$ & $17.97 \pm 0.01$ \\
		2021 Mar 21 & 2459294.79 & $17.43 \pm 0.03$ & $17.59 \pm 0.01$ & $17.55 \pm 0.01$ & $17.66 \pm 0.01$ & $17.72 \pm 0.02$ & $17.93 \pm 0.01$ \\
		2021 Mar 22 & 2459295.66 & $17.76 \pm 0.03$ & $17.74 \pm 0.01$ & $17.64 \pm 0.01$ & $17.62 \pm 0.02$ & $17.72 \pm 0.01$ & $17.81 \pm 0.01$ \\
		2021 Mar 22 & 2459295.82 & $17.89 \pm 0.03$ & $17.86 \pm 0.01$ & $17.71 \pm 0.01$ & $17.69 \pm 0.01$ & $17.70 \pm 0.02$ & $17.84 \pm 0.01$ \\
		2021 Mar 23 & 2459296.60 & $18.48 \pm 0.04$ & $18.15 \pm 0.01$ & $17.99 \pm 0.01$ & $17.91 \pm 0.03$ & $17.87 \pm 0.01$ & $17.87 \pm 0.01$ \\
		2021 Mar 23 & 2459296.78 & $18.58 \pm 0.03$ & $18.22 \pm 0.01$ & $18.05 \pm 0.01$ & $17.98 \pm 0.02$ & $17.89 \pm 0.01$ & $17.87 \pm 0.01$  \\
		2021 Mar 24 & 2459297.66 & $19.05 \pm 0.04$ & $18.50 \pm 0.01$ & $18.28 \pm 0.01$ & $18.14 \pm 0.01$ & $18.02 \pm 0.01$ & $17.98 \pm 0.01$  \\
		2021 Mar 24 & 2459297.78 & $19.05 \pm 0.04$ & $18.53 \pm 0.01$ & $18.32 \pm 0.01$ & $18.12 \pm 0.01$ & $18.04 \pm 0.01$ & $17.97 \pm 0.01$  \\
		2021 Mar 26 & 2459299.64 & $19.24 \pm 0.07$ & $18.39 \pm 0.02$ & $18.19 \pm 0.01$ & $17.97 \pm 0.01$ & $17.82 \pm 0.01$ & $17.82 \pm 0.01$  \\
		2021 Mar 26 & 2459299.78 & $19.12 \pm 0.06$ & $18.44 \pm 0.02$ & $18.16 \pm 0.01$ & $17.98 \pm 0.01$ & $17.79 \pm 0.01$ & $17.80 \pm 0.01$ \\
		2021 Mar 27 & 2459300.73 & $19.01 \pm 0.08$ & $18.27 \pm 0.02$ & $18.02 \pm 0.02$ & $17.78 \pm 0.01$ & $17.67 \pm 0.01$ & $17.61 \pm 0.01$  \\
		2021 Mar 31 & 2459304.71 & $18.62 \pm 0.05$ & $17.89 \pm 0.01$ & $17.59 \pm 0.01$ & $17.36 \pm 0.01$ & $17.21 \pm 0.01$ & $17.16 \pm 0.01$  \\
		2021 Apr 01 & 2459305.67 & $18.64 \pm 0.04$ & $17.88 \pm 0.01$ & $17.56 \pm 0.01$ & $17.39 \pm 0.02$ & $17.19 \pm 0.01$ & $17.11 \pm 0.01$  \\
		2021 Apr 02 & 2459306.67 & $18.69 \pm 0.04$ & $17.88 \pm 0.01$ & $17.57 \pm 0.01$ & $17.32 \pm 0.01$ & $17.12 \pm 0.01$ & $17.04 \pm 0.01$  \\
		2021 Apr 03 & 2459307.62 & $18.81 \pm 0.05$ & $17.96 \pm 0.01$ & $17.71 \pm 0.01$ & $17.41 \pm 0.01$ & $17.14 \pm 0.01$ & $17.03 \pm 0.01$  \\
		2021 Apr 05 & 2459309.69 & $19.32 \pm 0.04$ & $18.22 \pm 0.01$ & $17.86 \pm 0.01$ & $17.50 \pm 0.02$ & $17.28 \pm 0.01$ & $17.09 \pm 0.01$  \\
		2021 Apr 06 & 2459310.68 & $19.53 \pm 0.04$ & $18.32 \pm 0.01$ & $17.90 \pm 0.01$ & $17.51 \pm 0.01$ & $17.23 \pm 0.01$ & $17.01 \pm 0.01$  \\
		2021 Apr 07 & 2459311.64 & $19.88 \pm 0.05$ &  $18.57 \pm 0.01$ & $18.09 \pm 0.01$ & $17.65 \pm 0.01$ & $17.29 \pm 0.01$ & $17.08 \pm 0.01$  \\
		2021 Apr 08 & 2459312.69 & $ - $ & $ - $ & $ - $ & $ - $ & $17.40 \pm 0.03$ & $ - $ \\
		2021 Apr 09 & 2459313.66 & $20.39 \pm 0.07$ & $19.01 \pm 0.01$ & $18.50 \pm 0.01$ & $17.98 \pm 0.01$ & $17.54 \pm 0.01$ & $17.08 \pm 0.01$ \\
		2021 Apr 10 & 2459314.67 & $20.55 \pm 0.07$ & $19.01 \pm 0.01$ & $18.59 \pm 0.01$ & $18.03 \pm 0.02$ & $17.53 \pm 0.01$ & $17.29 \pm 0.01$  \\
		2021 Apr 11 & 2459315.62 & $20.86 \pm 0.10$ & $19.33 \pm 0.02$ & $18.78 \pm 0.01$ & $18.16 \pm 0.03$ & $17.65 \pm 0.01$ & $17.39 \pm 0.01$  \\
		2021 Apr 12 & 2459316.62 & $21.01 \pm 0.10$  & $19.42 \pm 0.02$ & $18.86 \pm 0.01$ & $18.25 \pm 0.03$ & $17.73 \pm 0.01$ & $17.45 \pm 0.01$  \\
		\hline
	\end{tabular}
\end{table*}

\begin{table*}
	\caption{Optical photometry of SN 2021gno with 0.8m Telescope at Baja and Konkoly Observatories}
	\label{tab:phot_opt_bajakonk}
	\begin{tabular}{lccccccc} 
		\hline
		Date & JD & $B$ & $g$ & $V$ & $r$ & $i$ & $z$ \\
		\hline
		2021 Mar 20 & 2459294.46 & $17.81 \pm 0.12$ & $17.53 \pm 0.02$ & $17.59 \pm 0.03$ & $17.76 \pm 0.02$ & $17.99 \pm 0.02$ & $18.38 \pm 0.07$ \\
		2021 Mar 21 & 2459294.51 & $17.87 \pm 0.12$ & $17.49 \pm 0.05$ & $17.56 \pm 0.06$ & $17.74 \pm 0.04$ & $17.89 \pm 0.06$ & $18.68 \pm 0.32$ \\
		2021 Mar 22 & 2459296.34 & $18.23 \pm 0.17$ & $18.13 \pm 0.09$ & $17.91 \pm 0.07$ & $17.91 \pm 0.04$ & $17.70 \pm 0.06$ & $18.16 \pm 0.23$ \\
		2021 Mar 25 & 2459299.46 & $-$ & $-$ & $18.17 \pm 0.16$ & $18.29 \pm 0.13$ & $17.50 \pm 0.07$ & $17.87 \pm 0.13$ \\
		2021 Mar 26 & 2459300.40 & $18.29 \pm 0.12$ & $18.10 \pm 0.07$ & $17.84 \pm 0.06$ & $17.93 \pm 0.05$ & $17.64 \pm 0.07$ & $18.59 \pm 0.25$ \\
		2021 Apr 01 & 2459305.55 & $18.03 \pm 0.08$ & $17.52 \pm 0.04$ & $17.27 \pm 0.04$ & $17.21 \pm 0.03$ & $17.00 \pm 0.02$ & $16.99 \pm 0.07$ \\
		2021 Apr 03 & 2459308.41 & $18.21 \pm 0.12$ & $17.60 \pm 0.03$ & $17.27 \pm 0.03$ & $17.17 \pm 0.03$ & $17.04 \pm 0.01$ & $17.24 \pm 0.03$ \\
		2021 Apr 04 & 2459309.41 & $18.31 \pm 0.11$ & $17.70 \pm 0.02$ & $17.39 \pm 0.04$ & $17.16 \pm 0.03$ & $17.05 \pm 0.02$ & $17.19 \pm 0.03$ \\
		2021 Apr 09 & 2459313.5 & $19.07 \pm 0.14$ & $18.32 \pm 0.03$ & $17.85 \pm 0.03$ & $17.49 \pm 0.02$ & $17.26 \pm 0.01$ & $17.38 \pm 0.04$ \\
		2021 Apr 10 & 2459315.44 & $19.15 \pm 0.12$ & $18.45 \pm 0.05$ & $18.16 \pm 0.06$ & $17.70 \pm 0.03$ & $17.39 \pm 0.04$ & $17.46 \pm 0.08$ \\
		2021 Apr 21 & 2459325.50 & $19.28 \pm 0.33$ & $19.58 \pm 0.31$ & $19.02 \pm 0.26$ & $18.43 \pm 0.11$ & $18.47 \pm 0.09$ & $-$ \\
		2021 Apr 24 & 2459328.31 & $19.45 \pm 0.23$ & $-$ & $18.68 \pm 0.12$ & $18.47 \pm 0.09$ & $18.49 \pm 0.18$ & $17.70 \pm 0.17$ \\
		2021 Jun 13 & 245937933 & $21.43 \pm 0.80$ & $21.97 \pm 0.55$ & $22.27 \pm 1.00$ & $20.21 \pm 0.17$ & $19.51 \pm 0.08$ & $19.76 \pm 0.27$ \\
		2021 Jun 26 & 2459392.33 & $20.36 \pm 0.21$ & $21.20 \pm 0.36$ & $20.63 \pm 0.23$ & $22.59 \pm 1.03$ & $19.60 \pm 0.09$ & $-$\\
	    \hline
	\end{tabular}
\end{table*}

\begin{table*}
	\caption{Optical photometry of SN 2021gno with LCOGT}
	\label{tab:phot_opt_lcogt}
	\begin{tabular}{lcccccc} 
		\hline
		Date & JD & $B$ & $g$ & $V$ & $r$ & $i$ \\
		\hline
		2021 Mar 24 & 2459297.61  & $18.44 \pm 0.03$ & $-$ & $ - $ & $ - $ & $-$  \\
		2021 Mar 26 & 2459299.52  & $18.41 \pm 0.04$ & $18.24 \pm 0.03$ & $ - $ & $ - $ & $-$  \\
		2021 Mar 31 & 2459305.32  & $17.84 \pm 0.02$ & $17.56 \pm 0.02$ & $ - $ & $ - $ & $-$  \\
		2021 Apr 02 & 2459307.28  & $17.90 \pm 0.02$ & $17.60 \pm 0.01$ & $ - $ & $ - $ & $-$  \\
		2021 Apr 14 & 2459319.37  & $19.78 \pm 0.10$ & $-$ & $ - $ & $ - $ & $-$  \\
		2021 Apr 16 & 2459321.22  & $19.97 \pm 0.17$ & $-$ & $ - $ & $ - $ & $-$  \\
        2021 Apr 28 & 2459332.96  & $20.20 \pm 0.12$ & $19.76 \pm 0.08$ & $ - $ & $ - $ & $18.30 \pm 0.03$  \\
		2021 May 03 & 2459337.93  & $20.54 \pm 0.07$ & $ - $ & $19.35 \pm 0.06$ & $ - $ & $18.41 \pm 0.03$  \\
		2021 May 08 & 2459343.25 & $20.82 \pm 0.12$ & $20.24 \pm 0.05$ & $19.70 \pm 0.05$ & $19.25 \pm 0.04$ & $18.62 \pm 0.04$  \\
		2021 May 10 & 2459345.23 & $20.88 \pm 0.10$ & $20.30 \pm 0.04$ & $19.70 \pm 0.04$ & $19.32 \pm 0.03$ & $18.62 \pm 0.03$  \\
		2021 May 13 & 2459348.93  & $20.77 \pm 0.09$ & $20.39 \pm 0.04$ & $19.92 \pm 0.04$ & $19.46 \pm 0.03$ & $18.80 \pm 0.03$  \\
		2021 May 18 & 2459353.38  & $20.98 \pm 0.17$ & $20.59 \pm 0.10$ & $20.07 \pm 0.08$ & $19.79 \pm 0.08$ & $18.93 \pm 0.07$  \\
		2021 May 23 & 2459358.48  & $21.18 \pm 0.60$ & $ - $ & $20.14 \pm 0.19$ & $ - $ & $19.00 \pm 0.07$  \\
		2021 May 27 & 2459361.57  & $20.96 \pm 0.29$ & $ - $ & $20.33 \pm 0.16$ & $ - $ & $19.06 \pm 0.05 $  \\
		2021 May 28 & 2459362.87  & $21.08 \pm 0.17$ & $20.86 \pm 0.10$ & $20.66 \pm 0.13$ & $20.16 \pm 0.04$ & $18.98 \pm 0.06$ \\
		2021 May 29 & 2459363.89   & $ - $ & $20.95 \pm 0.07$ & $ - $ & $20.10 \pm 0.03$ & $19.04 \pm 0.01$ \\
		2021 May 30 & 2459364.57 & $20.84 \pm 0.20$ & $ - $ & $20.55 \pm 0.11$ & $ - $ & $19.11 \pm 0.03$  \\
		2021 May 30 & 2459365.30  & $21.60 \pm 0.20 $ & $ - $ & $20.51 \pm 0.08$ & $ - $ & $19.15 \pm 0.04$  \\
		2021 Jun 02 & 2459367.57  & $21.67 \pm 0.17$ & $ - $ & $20.53 \pm 0.07$ & $ - $ & $19.19 \pm 0.03$ \\
		2021 Jun 05 & 2459370.68 & $ - $ & $ - $ & $21.02 \pm 0.79$ & $ - $ & $18.95 \pm 0.14$ \\
    	\hline
	\end{tabular}
\end{table*}
  
\begin{table*}
	\caption{Swift-UVOT photometry of SN 2021gno}
	\label{tab:phot_uv}
	\begin{tabular}{lccccccc} 
		\hline
		Date & JD & $w2$ & $m2$ & $w1$ & $u$ & $b$ & $v$\\
		\hline
		2021 Mar 20 & 2459294.044 & $15.21 \pm 0.06$ & $15.27 \pm 0.06$ & $15.58 \pm 0.06$ & $16.04 \pm 0.07$ & $17.36 \pm 0.12$ & $17.58 \pm 0.24$ \\
		2021 Mar 20 & 2459294.379 & $15.47 \pm 0.06$ & $15.44 \pm 0.05$ & $15.64 \pm 0.06$ & $16.09 \pm 0.07$ & $17.23 \pm 0.11$ & $17.71 \pm 0.27$ \\
		2021 Mar 21 & 2459294.940 & $16.04 \pm 0.06$ & $15.94 \pm 0.06$ & $16.08 \pm 0.07$ & $16.27 \pm 0.07$ & $17.25 \pm 0.11 $ & $17.65 \pm 0.26$ \\
		2021 Mar 22 & 2459296.010 & $17.26 \pm 0.08$ & $17.06 \pm 0.08$ & $16.97 \pm 0.09$ & $16.80 \pm 0.10$ & $17.62 \pm 0.15 $ & $17.50 \pm 0.22$ \\
		2021 Mar 23 & 2459297.096 & $18.48 \pm 0.13$ & $18.68 \pm 0.18$ & $18.24 \pm 0.21$ & $17.76 \pm 0.20$ & $18.16 \pm 0.24$ & $ - $ \\
		2021 Mar 25 & 2459298.783 & $19.21 \pm 0.22$ & $ - $ & $ - $ & $18.24 \pm 0.30$ & $18.20 \pm 0.26$ & $ - $ \\
		2021 Mar 29 & 2459303.165 & $19.13 \pm 0.20$ & $ - $ & $ - $ & $17.52 \pm 0.18$ & $17.62 \pm 0.16$ & $17.70 \pm 0.30$ \\
		2021 Mar 30 & 2459303.643 & $19.17 \pm 0.22$ & $ - $ & $ - $ & $17.70 \pm 0.21$ & $17.64 \pm 0.17$ & $17.07 \pm 0.19$ \\
		2021 Apr 04 & 2459309.216 & $19.58 \pm 0.26$ & $ - $ & $ - $ & $18.00 \pm 0.25$ & $17.82 \pm 0.19$ & $17.25 \pm 0.21$ \\
		2021 Apr 21 & 2459326.145 & $19.63 \pm 0.33$ & $ - $ & $ - $ & $ - $ & $ - $ & $ - $ \\
    	\hline
	\end{tabular}
\end{table*}
\newcommand{\angstrom}{\mbox{\normalfont\AA}}
\begin{table*}
	\caption{Bolometric luminosity of SN 2021gno}
	\label{tab:lbol}
	\begin{tabular}{lcccccc} 
		\hline
		Date & JD & Phase & $F_{u\rightarrow z}$ & $F_{qbol}$& $F_{bol}$ & $L_{bol}$ \\
		&&[days]& $[\times 10^{-12} \mathrm{erg~s^{-1}~cm^{-2}}$$\angstrom$$^{-1}]$ & $[\times 10^{-12} \mathrm{erg~s^{-1}~cm^{-2}}$$\angstrom$$^{-1}]$  & $[\times 10^{-12} \mathrm{erg~s^{-1}~cm^{-2}}$$\angstrom$$^{-1}]$ & $[\times 10^{41}\mathrm{erg~s^{-1}}]$ \\
		\hline
2021 Mar 20 & 2459294.04 & 1.24 & $ - $ & $ 5.90 \pm 0.18$ & $ 18.6 \pm 4.78 $ & $34.69 \pm 8.91$  \\
2021 Mar 20 & 2459294.37 & 1.57 & $ - $ & $ 5.47 \pm 0.17$ & $ 13.0 \pm 2.23 $ & $24.24 \pm 4.16$ \\
2021 Mar 20 & 2459294.46 & 1.66 & $ - $ & $ 5.83 \pm 0.14$ & $ 12.0 \pm 0.41 $ & $22.49 \pm 0.76$\\
2021 Mar 21 & 2459294.51 & 1.71 & $ - $& $ 5.62 \pm 0.12$ & $ 10.7 \pm 0.30 $ & $20.04 \pm 0.55$ \\
2021 Mar 21 & 2459294.6  & 1.80 & $1.87 \pm 0.02$ & $ 3.54 \pm 0.08$ & $ 8.20 \pm 0.37 $ & $15.28 \pm 0.69$ \\
2021 Mar 21 & 2459294.79 & 1.99 & $1.81 \pm 0.02$ & $ 5.05 \pm 0.11$ & $ 9.29 \pm 0.13 $ & $17.29 \pm 0.24$ \\
2021 Mar 21 & 2459294.94 & 2.14 & $1.81 \pm 0.02$ & $ 4.63 \pm 0.10$ & $ 7.00 \pm 0.24 $ & $13.04 \pm 0.45$ \\
2021 Mar 22 & 2459295.66 & 2.85 & $1.66 \pm 0.01$ & $ 3.36 \pm 0.07$ & $ 4.31 \pm 0.07 $ & $8.02\pm 0.14$ \\
2021 Mar 22 & 2459295.82 & 3.01 & $1.55 \pm 0.01$ & $ 2.99 \pm 0.05$ & $ 3.79 \pm 0.06 $ & $7.07 \pm 0.11$ \\
2021 Mar 22 & 2459296.01 & 3.20 & $1.48 \pm 0.02$ & $ 2.77 \pm 0.07$ & $ 3.39 \pm 0.08 $ & $6.32 \pm 0.15$\\
2021 Mar 22 & 2459296.34 & 3.53 & $1.26 \pm 0.04$ & $ 2.18 \pm 0.04$ & $ 2.72 \pm 0.05 $ & $5.07 \pm 0.10$ \\
2021 Mar 23 & 2459296.6  & 3.79 & $1.23 \pm 0.01$  & $ 1.88 \pm 0.03$ & $ 2.33 \pm 0.03 $ & $4.33 \pm 0.06$ \\
2021 Mar 23 & 2459296.78 & 3.96 & $1.18 \pm 0.01$ & $ 1.76 \pm 0.02$ & $ 2.19 \pm 0.03 $ & $4.07 \pm 0.05$\\
2021 Mar 23 & 2459297.09 & 4.27 & $1.11 \pm 0.01$ & $ 1.51 \pm 0.05$ & $ 1.93 \pm 0.05 $ & $3.60 \pm 0.10$ \\
2021 Mar 24 & 2459297.66 & 4.84 & $0.97 \pm 0.01$ & $ 1.32 \pm 0.02$ & $ 1.71 \pm 0.03 $ & $3.19 \pm 0.05$\\
2021 Mar 24 & 2459297.78 & 4.96 & $0.97 \pm 0.01$ & $ 1.30 \pm 0.02$ & $ 1.70 \pm 0.03 $ & $3.17 \pm 0.05$\\
2021 Mar 25 & 2459298.78 & 5.95 & $0.97 \pm 0.02$ & $ 1.26 \pm 0.06$ & $ 1.68 \pm 0.08 $ & $3.13 \pm 0.15$\\
2021 Mar 25 & 2459299.46 & 6.63 & $ - $ & $ 1.21 \pm 0.05$ & $ 1.90 \pm 0.28 $ & $3.55 \pm	0.52$\\
2021 Mar 26 & 2459299.64 & 6.81 & $1.09 \pm 0.01$ & $ 1.34 \pm 0.02$ & $ 1.83 \pm 0.03 $ & $3.40 \pm 0.06$ \\
2021 Mar 26 & 2459299.78 & 6.95 & $1.11 \pm 0.01$ & $ 1.38 \pm 0.02$ & $ 1.88 \pm 0.03 $ & $3.40 \pm 0.06$\\
2021 Mar 26 & 2459300.4  & 7.56 & $1.15 \pm 0.03$ & $ 1.43 \pm 0.04$ & $ 1.84 \pm 0.09 $ & $3.42 \pm 0.18$\\
2021 Mar 27 & 2459300.73 & 7.89 & $1.28 \pm 0.02$ & $ 1.57 \pm 0.03$ & $ 2.16 \pm 0.04 $ & $4.03 \pm 0.08$ \\
2021 Mar 29 & 2459303.16 & 10.31 & $1.71 \pm 0.07$ & $ 2.16 \pm 0.09$ & $ 2.91 \pm 0.21 $ & $5.42 \pm 0.39$\\
2021 Mar 30 & 2459303.64 & 10.78 & $1.78 \pm 0.06$ & $ 2.24 \pm 0.09$ & $ 3.06 \pm 0.19 $ & $5.70 \pm 0.36$ \\
2021 Mar 31 & 2459304.71 & 11.85 & $ 1.91 \pm 0.03$ & $ 1.91 \pm 0.03$ & $ 3.09 \pm 0.04 $ & $5.76 \pm 0.08$\\
2021 Apr 1 & 2459305.55  & 12.68 & $ 2.01 \pm 0.03$ & $ 2.01 \pm 0.03$ & $ 3.39 \pm 0.11 $ & $6.32 \pm 0.20$\\
2021 Apr 1 & 2459305.67  & 12.80 & $ 1.93 \pm 0.03$ & $ 1.93 \pm 0.03$ & $ 3.17 \pm 0.04 $ & $5.91 \pm 0.08$\\
2021 Apr 2 & 2459306.67  & 13.79 & $ 2.01 \pm 0.03$ & $ 2.01 \pm 0.03$ & $ 3.33 \pm 0.04 $ & $6.21 \pm 0.08$\\
2021 Apr 3 & 2459307.62  & 14.74 & $ 1.91 \pm 0.03$ & $ 1.91 \pm 0.03$ & $ 3.28 \pm 0.04 $ & $6.11 \pm 0.08$\\
2021 Apr 3 & 2459308.41  & 15.52 & $ 1.89 \pm 0.03$ & $ 1.89 \pm 0.03$ & $ 3.08 \pm 0.07 $ & $5.73 \pm 0.14$\\
2021 Apr 4 & 2459309.21  & 16.32 & $ 1.77 \pm 0.05$ & $ 1.98 \pm 0.12$ & $ 3.01 \pm 0.38 $ & $5.61 \pm 0.71$\\
2021 Apr 4 & 2459309.41  & 16.52 & $1.80 \pm 0.03$ & $ 1.80 \pm 0.03$ & $ 3.00 \pm 0.07 $ & $5.60 \pm  0.13$ \\
2021 Apr 5 & 2459309.69  & 16.79 & $1.68 \pm 0.03$ & $ 1.68 \pm 0.03$ & $ 3.00 \pm 0.04 $ & $5.59 \pm 0.08$\\
2021 Apr 6 & 2459310.68  & 17.78 & $1.68 \pm 0.03$ & $ 1.68 \pm 0.03$ & $ 3.15 \pm 0.04 $ & $5.87 \pm 0.09$ \\
2021 Apr 7 & 2459311.64  & 18.73 & $1.51 \pm 0.03$ & $ 1.51 \pm 0.03$ & $ 2.96 \pm 0.04 $ & $5.51 \pm  0.08$\\
2021 Apr 8 & 2459312.69  & 19.78 & $1.32 \pm 0.05$ & $ 1.30 \pm 0.07$ & $ 3.10 \pm 0.29 $ & $5.77 \pm 0.55$\\
2021 Apr 9 & 2459313.5   & 20.58 & $1.22 \pm 0.01$ & $ 1.22 \pm 0.01$ & $ 2.38 \pm 0.07 $ & $4.43 \pm 0.13$\\
2021 Apr 9 & 2459313.66  & 20.74 & $1.23 \pm 0.03$ & $ 1.23 \pm 0.03$ & $ 3.00 \pm 0.05 $ & $5.59 \pm 0.09$\\
2021 Apr 10 & 2459314.67 & 21.74 & $1.15 \pm 0.03$ & $ 1.15 \pm 0.03$ & $ 2.58 \pm 0.04 $ & $4.80 \pm 0.08$\\
2021 Apr 10 & 2459315.44 & 22.51 & $1.04 \pm 0.02$ & $ 1.04 \pm 0.02$ & $ 2.12 \pm 0.12 $ & $3.95 \pm 0.23$\\
2021 Apr 11 & 2459315.62 & 22.69 & $1.01 \pm 0.03$ & $ 1.01 \pm 0.03$ & $ 2.40 \pm 0.04 $ & $4.47 \pm 0.08$\\
2021 Apr 12 & 2459316.62 & 23.68 & $0.94 \pm 0.03$ & $ 0.94 \pm 0.03$ & $ 2.27 \pm 0.04 $ & $4.24 \pm 0.08$ \\
2021 Apr 21 & 2459325.5  & 32.51 & $ - $ & $ 0.43 \pm 0.03$ & $ 0.90 \pm 0.15 $ & $1.74 \pm	0.28$\\
2021 Apr 24 & 2459328.31 & 35.30 & $ - $ & $ 0.46 \pm 0.02$ & $ 0.91 \pm 0.12 $ & $1.77 \pm	0.23$\\
2021 Apr 28 & 2459332.96 & 39.92 & $ - $ & $ 0.22 \pm 0.01$ & $ 0.74 \pm 0.07 $ & $1.44 \pm	0.13$\\
2021 May 3 & 2459337.93  & 44.86 & $ - $ & $ 0.20 \pm 0.01$ & $ 0.63 \pm 0.04 $ & $1.23 \pm 0.08$\\
2021 May 8 & 2459343.25  & 50.15 & $ - $ & $ 0.16 \pm 0.01$ & $ 0.43 \pm 0.03 $ & $0.85 \pm	0.06$ \\
2021 May 10 & 2459345.23 & 52.12 & $ - $ & $ 0.15 \pm 0.01$ & $ 0.47 \pm 0.03 $ & $0.91 \pm	0.05$ \\
2021 May 13 & 2459348.93 & 55.79 & $ - $ & $ 0.13 \pm 0.01$ & $ 0.38 \pm 0.02 $ & $0.74 \pm	0.04$\\
2021 May 18 & 2459353.38 & 60.22 & $ - $ & $ 0.11 \pm 0.01$ & $ 0.32 \pm 0.04 $ & $0.62 \pm	0.07$\\
2021 May 23 & 2459358.48 & 65.28 & $ - $ & $ 0.10 \pm 0.01$ & $ 0.28 \pm 0.08 $ & $0.58 \pm	0.15$\\
2021 May 27 & 2459361.57 & 68.36 & $ - $ & $ 0.09 \pm 0.01$ & $ 0.27 \pm 0.05 $ & $0.55 \pm	0.10$\\
2021 May 28 & 2459362.87 & 69.65 & $ - $ & $ 0.08 \pm 0.01$ & $ 0.28 \pm 0.04 $ & $0.55 \pm	0.07$\\
2021 May 30 & 2459364.57 & 71.34 & $ - $ & $ 0.08 \pm 0.01$ & $ 0.27 \pm 0.04 $ & $0.55 \pm	0.07$\\
2021 May 30 & 2459365.3  & 72.06 & $ - $ & $ 0.07 \pm 0.01$ & $ 0.25 \pm 0.03 $ & $0.50 \pm	0.06$\\
2021 Jun 2 & 2459367.57  & 74.32 & $ - $ & $ 0.07 \pm 0.01$ & $ 0.24 \pm 0.02 $ & $0.49 \pm	0.05$ \\
    	\hline
	\end{tabular}
\end{table*}

%%%%%%%%%%%%%%%%%%%%%%%%%%%%%%%%%%%%%%%%%%%%%%%%%%

% Don't change these lines
\bsp	% typesetting comment
\label{lastpage}
\end{document}